\documentclass[acmsmall,screen,review=false]{acmart}

\usepackage{graphicx}
\usepackage{amsmath}
\usepackage{colortbl}
\usepackage{arydshln}
\usepackage{multirow}
\usepackage{enumerate}
\usepackage{makecell}
\usepackage{amsmath,amsfonts}
\usepackage{algorithmic}
\usepackage{graphicx}
\usepackage{textcomp}
\usepackage{xcolor}
\usepackage{threeparttable}
\usepackage{caption}
\usepackage{varioref}
\usepackage{cleveref}
\usepackage{booktabs}
\usepackage[most]{tcolorbox}
\definecolor{lightgray}{RGB}{249, 249, 249}

\usepackage{adjustbox}
\newtcolorbox{mybox}{
  breakable,
  arc=3mm,
  colback=lightgray,
  colframe=black,
  boxrule=0.3mm,
}

\AtBeginDocument{
  }

\setcopyright{acmcopyright}
\copyrightyear{2023}
\acmYear{2023}
\acmDOI{XXXXXXX.XXXXXXX}
\acmJournal{TOSEM}

\begin{document}

\title{Self-planning Code Generation with Large Language Models}

\author{Xue Jiang}
\email{jiangxue@stu.pku.edu.cn}
\author{Yihong Dong}
\email{dongyh@stu.pku.edu.cn}
\author{Lecheng Wang}
\email{wanglecheng@stu.pku.edu.cn}
\author{Zheng Fang}
\email{fangz@pku.edu.cn}
\author{Qiwei Shang}
\email{shangqiwei@stu.pku.edu.cn}
\author{Ge Li}
\email{lige@pku.edu.cn}
\authornote{Corresponding author}
\author{Zhi Jin}
\email{zhijin@pku.edu.cn}
\author{Wenpin Jiao}
\email{jwp@sei.pku.edu.cn}
\affiliation{%
  \institution{Key Laboratory of High Confidence Software Technologies (Peking University), Ministry of Education; School of Computer Science, Peking University, Beijing}
  \country{China}
}

\renewcommand{\shortauthors}{Jiang et al.}

\begin{abstract}
Although large language models (LLMs) have demonstrated impressive ability in code generation, they are still struggling to address the complicated intent provided by humans. It is widely acknowledged that humans typically employ planning to decompose complex problems and schedule solution steps prior to implementation. To this end, we introduce planning into code generation to help the model understand complex intent and reduce the difficulty of problem-solving. This paper proposes a self-planning code generation approach with large language models, which consists of two phases, namely planning phase and implementation phase. 
Specifically, in the planning phase, LLM outlines concise and formatted planning steps from the intent. Subsequently, in the implementation phase, the model generates code step by step, guided by the preceding planning steps. 
We conduct extensive experiments on various code-generation benchmarks across multiple programming languages. Experimental results show that self-planning code generation achieves a relative improvement of up to 25.4\% in Pass@1 compared to direct code generation, 
and up to 11.9\% compared to Chain-of-Thought code generation.
Moreover, our self-planning approach also enhances the quality of the generated code with respect to correctness, readability, and robustness, as assessed by humans.
\end{abstract}

\ccsdesc[500]{Software and its engineering~Software creation and management}
\ccsdesc[500]{Computing methodologies~Artificial intelligence}

\maketitle

\newcommand{\chapquote}[3]{
\begin{list}{}{
    \setlength{\leftmargin}{15pt}
    \setlength{\rightmargin}{0pt}
  }
  \item \textit{#1} 
  \begin{flushright} - #2 \end{flushright}
\end{list}}
\chapquote{``The art of programming is the art of organizing complexity."}{Edsger W.Dijkstra}

\section{Introduction}
Programming is a pervasive and powerful tool for problem-solving. As one of the most central problems in programming theory, code generation allows machines to program automatically to satisfy human intent expressed in the form of some specification. 
In recent years, code generation has achieved great progress in both academia and industry \cite{Synchromesh, CODEP, skcoder, Subtoken-TranX, codet}. In particular, LLMs \cite{GPT3,codex} demonstrate impressive code generation abilities, attracting attention from various fields such as artificial intelligence, natural language processing (NLP), and software engineering.

In code generation, the human-provided intent is usually a natural language description of "what to do" problem, while the model solves the problem by generating "how to do" code. When the intent is straightforward, it is easy to map to the code, which can be well handled by state-of-the-art code generation models \cite{codex, alphacode}. However, as the problem becomes complicated and scaled, directly generating complex code satisfying intent is challenging for both people and models (even LLMs). In practice, software development is to give software solutions for real-world problems, and the generation of these solutions requires a planning process to guarantee the quality of coding \cite{abrahamsson2002agile, Waterfall, lifecycle_models}. Accordingly, programmers outline a plan in advance and then complete the entire code step by step following the plan. For complex code generation tasks, such planning is not just beneficial, it is imperative. Therefore, we desire to incorporate planning into code generation. Plan-aided code generation has the following two benefits. 1) It breaks down the complex problem into several easy-to-solve subproblems, which reduces the difficulty of problem-solving. 2) It abstracts the problem and provides instructions for solving it, which helps the model understand how to generate code. Therefore, planning in advance can facilitate the generation of correct codes for complex problems. 

Generally, plan-aided code generation presupposes the existence of an approach for converting intent into plan. However, if we build such an approach from scratch, it requires a large amount of resources to label intent-plan pairs for training a model. Few-shot prompting provides an important way of using LLMs without training. A successful technique of few-shot prompting is Chain of Thought (CoT) \cite{wei2022chain}, which enables LLMs to perform step-by-step reasoning to solve reasoning tasks, such as mathematical \cite{mathreason}, commonsense \cite{TalmorTCGB20}, and symbolic reasoning \cite{YaoZYDSN023}. The ability to generate CoTs, as demonstrated by the LLMs, helps us to achieve planning. Nonetheless, applying CoT in the process of planning for code generation remains challenging. Fundamentally, CoT and code \footnote{The program is defined as a collection of commands written in a computer language to achieve a specific goal or solve a specific problem.} are both descriptions of solutions to achieve the final goal (i.e., output), just in different forms: one in natural language and the other in programming language\footnote{The authors of CoT also discussed the relationship between CoT and program synthesis and execution in extended related work \cite{wei2022chain}, considering their work as a generalization of program synthesis and execution in the natural language domain.}. Generating solutions to problems, whether through CoT or code, presents similar challenges. Using CoT directly does not reduce the difficulty of code generation basically. Consequently, we implement planning based on the principle of problem decomposition.

In this paper, we propose a self-planning code generation approach with LLMs that exploits the planning capabilities of LLMs themselves to facilitate code generation. Self-planning code generation consists of two phases during inference:
1) Planning phase, LLM generates plans for problems by providing only a few intent-to-plan demonstrations as examples in prompting;
2) Implementation phase, the LLM generates code that adheres to the intent step by step, guided by the plans.
Self-planning code generation leverages few-shot prompting to generate plans autonomously without annotating plan corpus and extra training. 
\textit{To the best of our knowledge, this work is the first to introduce CoT to code generation (i.e., Code CoT)\footnote{Detailed Illustrations and Discussions can be found in Fig. \ref{vCoT} and Fig. \ref{planCoT} of Appendix.} and propose a planning-based approach to address the uncovered inefficiencies of CoT.}

\begin{figure*}[t!]
\centering
\includegraphics[width=1\textwidth]{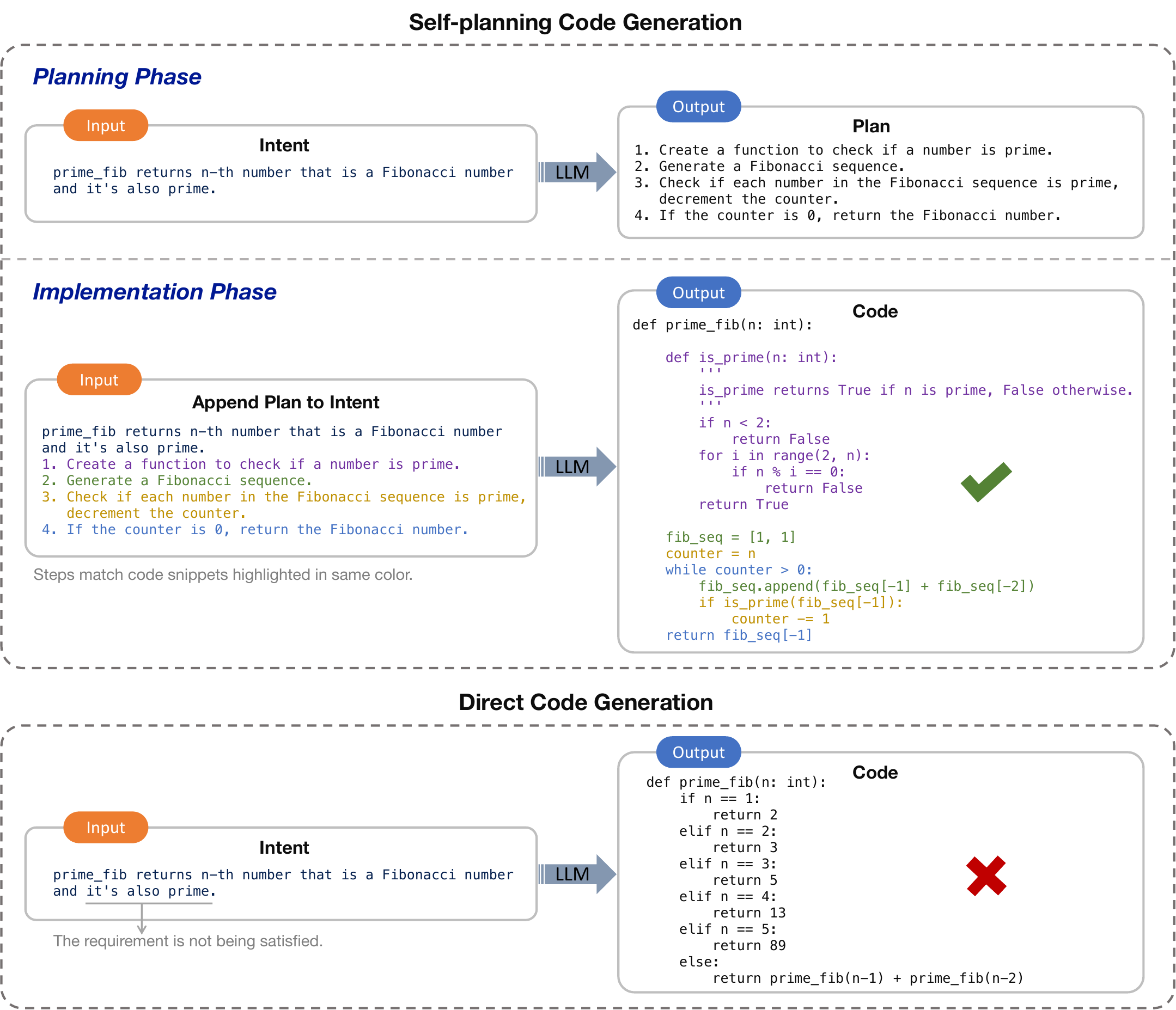}
\caption{Self-planning code generation is carried out in two phases (i.e., planning phase and implementation phase): 1) In planning phase, LLM decomposes an intent into a set of easy-to-solve sub-problems and devises a plan for executing the solution steps; 2) In implementation phase, LLM generates code following the intent and plan, which assists self-planning code generation to be capable of handling more difficult problems than direct code generation with LLM. Direct code generation uses the intent as input to the LLM and the LLM generates the code directly. The LLM used here to demonstrate the example of code generation is code-davinci-002.}
\label{self-plan}
\end{figure*}

Empirical evaluations have provided evidence that self-planning approach can substantially improve the performance of LLMs on code generation. 
1) Self-planning approach showed a relative improvement of up to 25.4\% in Pass@1 over the direct generation approach and up to 11.9\% over CoT approach of code generation.
2) We show that self-planning is an emergent ability that appears on large enough LLMs, but planning can benefit most LLMs.
3) We explore several variants of the self-planning approach in depth and demonstrate that our designed self-planning approach is the optimal choice in these variants.
4) We validate the effectiveness of self-planning approach across multiple programming languages (PLs) including Python, Java, Go, and JavaScript.
5) We analyze the quality (i.e., correctness, readability, and robustness) of the code generated by self-planning approach through human evaluation.

\section{Self-planning}
In self-planning code generation, we propose conducting planning prior to the actual code generation by LLMs. This process can be divided into two phases.

\textbf{Planning phase.} \, 
In the planning phase, we employ an LLM to abstract and decompose the intent to obtain a plan for guiding code generation. We take advantage of the ability of LLM to perform planning through few-shot prompting. 
In few-shot prompting, we require only a few labeled examples to demonstrate the task at hand as a prompt. Subsequently, this prompt is incorporated into the model input during inference, enabling the model to adhere to the prompt in order to accomplish the given task.

In our approach, the prompt $C$ is specifically designed as $k$ examples concatenated together, i.e., $C \triangleq \langle x^e_1 \cdot y^e_1\rangle\parallel\langle x^e_2 \cdot y^e_2\rangle\parallel...\parallel\langle x^e_k \cdot y^e_k\rangle$, where each example $\langle x^e_i \cdot y^e_i\rangle$ consists of the example intent $x^e_i$ and its associated plan $y^e_i$ to demonstrate the planning task. The plan is a scheduling of subproblems that abstract and decompose from intent, which is set to $y^e_i = \{q_{i,j}\}^n_{j=1}$, where $q_{i,j}$ is $j$-th step in plan $y^e_i$. During inference, the test-time intent $x$ will be concatenated after the prompt, and $C \parallel x$ will be fed into the LLM $\mathcal{M}$, which will attempt to do planning for the test-time intent. Thus, we can obtain the test-time plan $y$.

Note that $k$ of the prompt is a fairly low number, meaning we can achieve self-planning by labeling only a few examples demonstrating planning.

\textbf{Implementation phase.} \, 
In the implementation phase, we use the plan obtained during the planning phase to guide LLM in generating code.
We append the plan $y$ to the intent $x$ as input for the LLM $\mathcal{M}$. The LLM generates the final code $z$ by way of predicting the next token.

The above two phases can be formalized as the following equation.
\begin{equation}
\begin{aligned}
    \mathcal{P}\left(z|x,C\right) &= \sum_{\hat{y}} \mathcal{P}\left(z|\hat{y},x,C\right)\cdot\mathcal{P}\left(\hat{y}|x,C\right),\\
    & \propto \mathcal{P}\left(z|y,x,C\right)\cdot\mathcal{P}\left(y|x,C\right),
\end{aligned}
\end{equation}

where $\hat{y}$ is any of all possible plans, and $y$ denotes one of the plans generated by $\mathcal{M}$. In this paper, we adopt the plan with the highest probability as $y$. We further simplify $\mathcal{P}\left(z|y,x,C\right) = \mathcal{P}\left(z|y,x\right)$ via conditional independence assumptions, thus:

\begin{equation}
\begin{aligned}
    \mathcal{P}\left(z|x,C\right) \triangleq 
    \underbrace{\mathcal{P}\left(z|y,x\right)}_{\text{Implementation phase}}\cdot \quad
    \underbrace{\mathcal{P}\left(y|x,C\right)}_{\text{Planning phase}},
\end{aligned}
\end{equation}

\textbf{Crafting prompts for self-planning.} \ According to the methodology of few-shot prompting, we need to construct some examples as prompts to instruct the model for planning. Therefore, we prepare $k$ example intents and write plans for each intent following the subsequent principles.

\begin{enumerate}[1.]
\item The plan is organized in the form of a numbered list, where each item in the list represents a step.

\item Every step represents a single, easily implementable sub-task. These sub-tasks are formulated as imperative sentences that start with verbs, focusing on the action needed in each step. 

\item The steps should be written concisely and at a high level, avoiding overly detailed implementation specifics. A step like "Check if a number is prime" is more appropriate than a detailed implementation such as "If the number is less than 2, it's not prime. Check if the number is divisible by any number between 2 and n-1. If the number is not divisible by any number between 2 and n-1, it's prime". 

\item The execution of the plan happens sequentially, but the plan can incorporate conditional (if) and looping (loop) keywords for more complex structures. This allows for branching paths and loops as necessary while still maintaining the logical progression of the plan.

\end{enumerate}
Self-planning prompts can be freely written within these simple principles, so the crafting of the prompts is relatively straightforward and efficient. 

\textbf{Example.} \ An example of self-planning code generation derived from the real benchmark HumanEval is shown in Fig. \ref{self-plan}. 
In the planning phase, human provides an intent to \textit{find the n-th number that is a Fibonacci number and it's also prime} \footnote{The self-planning prompt is appended before the intent, guiding LLM to perform planning, which we've omitted in Fig. \ref{self-plan} for presentation purposes.}. LLM abstracts two subproblems from the intent, i.e., \textit{generating a Fibonacci sequence} and \textit{determining if a number is prime}, and plans four steps to solve the subproblems combinatorial. Then entering the implementation phase, we append the plan to the intent and feed it to LLM. LLM generates code under the navigation of the steps, and surprisingly, it wraps "\textit{determine if a number is prime}" into a subfunction and calls it. At the same time,  the previously generated plan significantly augments the readability of the code, which is an important aspect of measuring code quality.

In contrast, LLM (e.g. code-davinci-002) cannot understand that the intent is a combination of multiple problems in direct code generation. LLM knows to write something about "prime" and "Fibonacci", but actually, it generates a confusing code, i.e. it enumerates the first five correct samples \footnote{This is related to the fact that the benchmark HumanEval provided five public test cases as additional input, and the model copied them.} and then calculating the Fibonacci numbers, completely losing the requirement to determine \textit{whether it is a prime number}.

In short, when code generation tasks become complex, incorporating planning to handle the complexity becomes necessary.

\section{Evaluation}
We evaluate our self-planning approach by addressing the following research questions (RQs):
\begin{itemize}
    \item \textbf{RQ1}: How does self-planning approach perform in code generation compared to baseline approaches?
    \item \textbf{RQ2}: How does the self-planning approach perform based on different LLMs?
    \item \textbf{RQ3}: What is the optimal design for the self-planning approach?
   \item \textbf{RQ4}: How does self-planning approach perform in multilingual code generation?
   \item \textbf{RQ5}: How does the complexity of the problem affect self-planning?
\end{itemize}

\subsection{Experiment Setup}
\subsubsection{Benchmarks}
Following the previous work \cite{gpt4, incoder, nijkamp2022codegen, codegeex}, we adopt two public mainstream benchmarks, MBPP and HumanEval, along with their multilingual versions and extended test case versions, to evaluate the code generation ability of our self-planning approach and various baselines.

\textbf{MBPP-sanitized.} \cite{mbpp} \ This benchmark is a manually verified subset of MBPP (Mostly Basic Programming Problems), contains 427 crowd-sourced Python programming problems, covering programming fundamentals, standard library functionality, and more. Each problem consists of an NL description, a code solution, and 3 automated test cases. For MBPP-sanitized, the NL description is provided as input.

\textbf{HumanEval.} \cite{codex} is a set of 164 handwritten programming problems, proposed by OpenAI. Each problem includes a function signature, NL description, function body, and several unit tests, with an average of 7.7 tests per problem. For HumanEval, function signature, NL description, and public test cases are provided as input.

\textbf{HumanEval-X.} \cite{codegeex} is constructed based on HumanEval to better evaluate the multilingual capabilities of code generation models. HumanEval-X consists of 820 high-quality human-crafted data samples (each with test cases) in Java, JavaScript, Go, etc..

\textbf{MBPP-ET.} and \textbf{HumanEval-ET} \cite{CodeScore} are two public expanded versions of MBPP and HumanEval, each including over 100 additional test cases per task. This updated version includes edge test cases that enhance the soundness of code evaluation in comparison to the original benchmark.

\subsubsection{Metrics}
To assess the accuracy of the generated code, we employ two types of metrics: an execution-based metric, i.e. \textbf{Pass@k} and \textbf{AvgPassRatio}, and a match-based metric, i.e. \textbf{CodeBLEU} (Details of metrics can be found in Appendix \ref{Details of Metrics}). The execution-based metrics measure the functional correctness of the generated code through executing the given test cases, and the match-based metrics measure the similarity between the generated code and the given reference code.

\subsubsection{Basic Baselines}
We conduct various experiments, comparing multiple baselines to evaluate distinct aspects. Among these, three baselines—Direct, Code CoT, and Ground-truth Planning, serve as basic baselines in all experiments, highlighting the efficacy of our approach.

\textbf{Direct} generates code using LLMs in a zero-shot setting, implying only intent and no examples are available in the prompt. 

\textbf{Code CoT} generates a chain of thought for each question by using Code CoT prompt (described in Crafting Prompt) and then generates the corresponding code. This approach aligns with self-planning in that both adopt a two-stage generation approach. 

\textbf{Ground-truth Planning} is set to investigate the maximum potential of the planning approach in code generation, we directly supply the model with ground-truth plans to perform the implementation phase, skipping the planning phase. 

\subsubsection{Implementation Details}
The implementation details of our experiment are as follows.

\textbf{Crafting Prompt.}
We use a different prompt for problems in MBPP and in HumanEval in a fixed way. We employ a simple method to select the questions used to construct the prompts, i.e., random sampling with fixing a seed. Specifically, for HumanEval, we randomly sample 8 questions from HumanEval to create the prompt. For MBPP, given MBPP has an additional small training set, we identified four representative categories of problems: string processing, numerical calculations, number theory problems, data structure manipulations, and geometric computations, and 8 problems are randomly sampled from these categories. Prompts for our approach and all baselines are constructed utilizing the same problems.

For the self-planning prompt, we manually crafted plans for the problems. Self-planning prompts for HumanEval and  MBPP are listed in Appendix \ref{self_planning_prompt_for_humanEval}.
\textit{For the baseline code CoT, it is implemented by ourselves according to the original paper of CoT, as there is currently no CoT prompt designed for code generation. The way to create a Code CoT prompt is by providing ground-truth code of the 8 problems and then using the instruction "\textit{Generate detailed comments for this code.}" to enable LLMs to generate comments as intermediate steps. 
To avoid bias caused by errors in LLMs generation and inconsistencies in style, the generated Code CoT prompts are manually reviewed and adapted to the same numbered list format as the self-planning prompts.} The instance of Code CoT prompt can be found in Appendix \ref{example_baseline}.
The examples selected from the dataset for prompting will be excluded from the evaluation.

\textbf{Ground-truth Plan Generation and Validation.}
Labeling plans for all datasets is labor-intensive. To mitigate this concern, we utilize the existing ground-truth code to inversely generate a plan, which we then adopt as the ground-truth plan in our experiments. We adopt the few-shot prompting approach to implement this strategy. By reusing self-planning prompts, we construct the corresponding prompts $C^p \triangleq \langle x^e_1 \cdot c^e_1 \cdot y^e_1\rangle\parallel\langle x^e_2 \cdot c^e_2 \cdot y^e_2\rangle\parallel...\parallel\langle x^e_k \cdot c^e_k \cdot y^e_k\rangle$, where each example $\langle x^e_i \cdot c^e_i \cdot y^e_i\rangle$ consists of the example intent $x^e_i$, ground-truth code $c^e_i$ from dataset, plan $y^e_i$ to demonstrate the planning task. The instance of the prompt can be found in Appendix \ref{example_baseline}.

We manually validated the generated ground-truth plans on HumanEval dataset. The experimental results show that most of the generated ground-truth plans follow the principles of self-planning and satisfy the requirements completely, with only a very small number (about 3\%) having poorly described steps. Therefore, the generated ground-truth plans are relatively high-quality.

\textbf{Model Configuration and Evaluation.}
All basic baselines adopt code-davinci-002 as the base model and set the max generation length to 300 by default. We obtain only one plan in the planning phase by greedy search. For the matrics Pass@1, AvgPassRatio, and CodeBLEU, we use the greedy search setting with temperature 0 and top $p$ 1 to generate one code. For Pass@k (k$\geq$2), we generate 10 samples for each problem in benchmarks and set temperature to 0.8 and top $p$ to 0.95.

\section{Experimental Results}

\subsection{Comparison With Baselines (RQ1)}
\textbf{Evaluation.} \, 
We conduct a comparison between the self-planning approach and the following baselines, which comprise our main experimental result. First, we benchmark our approach against a range of widely recognized LLMs pre-trained on code, including AlphaCode (1.1B) \cite{alphacode}, Incoder (6.7B) \cite{incoder}, CodeGeeX (13B) \cite{codegeex}, CodeGen-Mono (16.1B) \cite{nijkamp2022codegen}, and PaLM Coder (560B) \cite{FlanPaLM}. The aim is to ascertain the performance level at which our approach operates relative to these recognized models. Second, we establish code-davinci-002 \cite{codex} as our base model and compare self-planning approach with Direct, Few-shot, and Code CoT to demonstrate the effectiveness of our approach, where the Few-shot approach uses the requirements and code pairs as prompt. Third, we investigate the impact of the ground-truth planning approach, which can be considered as an underestimated upper bound for the base model employing self-planning. Fourth, we sampled the code during LLM generation to investigate whether planning affected the diversity of the generated code. Note that sampling is limited to code rather than plan.

\begin{table*}[h!]
\setlength\tabcolsep{1pt}
\caption{Comparison of self-planning approaches and various baselines, and the number after $\uparrow$ denotes the performance improvement achieved in LLM upon incorporating the corresponding approach, i.e., the relative improvement compared to approach Direct.}
\centering
\resizebox{1\textwidth}{!}{
\begin{tabular}{lcccccccccc}
\toprule
\multicolumn{1}{l}{\multirow{2}{*}{Approach}} &
\multicolumn{3}{c}{\cellcolor{gray!10} HumanEval} &
\multicolumn{2}{c}{\cellcolor{gray!20} HumanEval-ET} &
\multicolumn{3}{c}{\cellcolor{gray!30} MBPP-sanitized} & 
\multicolumn{2}{c}{\cellcolor{gray!40} MBPP-ET} \\
\cmidrule(r){2-4} \cmidrule(r){5-6} \cmidrule(r){7-9} \cmidrule(r){10-11}
  \multicolumn{1}{l}{} & 
  \multicolumn{1}{c}{Pass@1} &
  \multicolumn{1}{l}{CodeBLEU} &
  \multicolumn{1}{c}{AvgPassRatio} &
  \multicolumn{1}{c}{Pass@1} &
  \multicolumn{1}{c}{AvgPassRatio} &
    \multicolumn{1}{c}{Pass@1} &
  \multicolumn{1}{l}{CodeBLEU} &
  \multicolumn{1}{c}{AvgPassRatio} &
  \multicolumn{1}{c}{Pass@1} &
  \multicolumn{1}{c}{AvgPassRatio}
 \\  
 \hline
\multicolumn{2}{l}{\textbf{Code pre-trained models}}\\
AlphaCode (1.1B) & 17.1 & - & - & - & -  & - & - & - & - & - \\
Incoder (6.7B) & 16.0 & 16.2 & 28.7 & 12.2 & 27.9 & 14.6 & 16.9 & 17.9 & 11.8 & 17.4 \\
CodeGeeX (13B) & 25.9 & 23.1 & 31.4 & 16.0 & 36.3 & 19.9 & 18.4 & 38.8 & 18.2 & 26.9\\
CodeGen(16.1B) & 34.6 & 22.8 & 57.5 & 26.3 & 52.6 & 36.6 & 24.5 & 41.6 & 28.1 & 36.9 \\
PaLM Coder (560B) & 36.0 & - & - & - & -  & - & - & - & - & - \\
\hdashline
Direct  & 48.1  & 24.0 & 63.2 & 37.2 & 62.7 & 49.8 & 25.6 & 54.8 & 37.7 & 46.4\\
Few-shot  & 52.6  & 28.2 & 74.9 & 44.2 & 72.8 & 53.5 & 26.1 & 56.9 & 38.2 & 48.8 \\
Code CoT  & $53.9 \ \ (\uparrow 12.1\%)$ & 30.4 & 75.6 & $45.5 \ \ (\uparrow 22.3\%)$ & 74.7 & $54.5 \ \ (\uparrow 9.4\%)$ & 26.4 & 58.7 & $39.6 \ \ (\uparrow 5.0\%)$ & 49.9\\
Self-planning  & $\textbf{60.3} \ \ (\textcolor{red}{\uparrow 25.4\%})$  & 28.6 & 80.8 &  $\textbf{46.2} \ \ (\textcolor{red}{\uparrow 24.1\%})$  & 76.4 & $\textbf{55.7} \ \ (\textcolor{red}{\uparrow 11.8\%})$  & 24.9 & 59.6 &  $\textbf{41.9} \ \ (\textcolor{red}{\uparrow 11.2\%})$ & 51.0\\
\hdashline
Ground-truth Planning  & $74.4 \ \ (\textcolor{teal}{\uparrow 54.7\%})$ & 41.0 & 88.1 & $57.7 \ \ (\textcolor{teal}{\uparrow 55.1\%})$ & 85.2 & $65.1 \ \ (\textcolor{teal}{\uparrow 30.7\%})$ & 33.7 & 69.0 & $50.7 \ \ (\textcolor{teal}{\uparrow 34.5\%})$ & 60.2\\

\bottomrule
\end{tabular}
}\label{main_results}
\end{table*}

\textbf{Results.} \, The results are summarized in Table \ref{main_results}, which demonstrate a significant effect of self-planning code generation. 
The self-planning approach is based on a powerful base model, which far outperforms other models pre-trained with code, even PaLM Coder, which has three times the number of parameters. The experimental results suggest that obtaining the plan or Code CoT from the intent can provide a noteworthy advantage in code generation compared to the direct generation of code from the intent. However, the advantage of Code CoT over Few-shot is marginal. Self-planning outperforms both Code CoT and Few-shot across four public code generation datasets, showing a notable improvement in Pass@1 over Code CoT and Few-shot. While our approach demonstrates slightly lower performance on the CodeBLEU metric, it's worth noting that CodeBLEU assesses the similarity between the generated and reference code. This metric can be limiting, as the reference code may not represent the sole valid solution --- a critique often associated with match-based metrics. Moreover, we evaluated the impact of utilizing the ground-truth plan in facilitating code generation. This approach simulates to some extent the ground-truth planning provided by developers and provides an understanding of the approximate upper bound (which is actually low) of the self-planning approach. The results in Table \ref{main_results} indicate a substantial improvement in the use of ground-truth planning, as evidenced by a relative improvement of over 50\% and 30\% on HumanEval and MBPP-sanitized benchmarks respectively. Overall, the self-planning approach showed a more significant improvement on HumanEval compared to on MBPP-sanitized. We hypothesize that this is due to the fact that in some of the MBPP-sanitized problems, the information provided about intentions is not sufficient to allow the model to perform an effective solution, and even humans are barely able to solve these problems.

Another result of Pass@k, with multiple samples, is shown in Table \ref{sample}. The diversity and accuracy of the self-planning approach consistently outperform Direct as the sample size increases. In contrast, the diversity of Code CoT decreases rapidly, and its Pass@5 and pass@10 are both lower than Direct, indicating that detailed solution steps entail a loss of diversity. Pass@k for ground-truth planning has been maintained at a high level. It is worth noting that when sampling 10 codes, Ground-truth planning is able to solve close to 90\% of tasks.

\begin{table}[h!]
\caption{Pass@k (\%) of self-planning and other approaches on HumanEval benchmarks.}
\centering
\small{
\begin{tabular}{lcccc}
\toprule
\multicolumn{1}{l}{\multirow{1}{*}{Approach}} & 
  \multicolumn{1}{c}{Pass@1} &
  \multicolumn{1}{c}{Pass@2} &
  \multicolumn{1}{c}{Pass@5} &
  \multicolumn{1}{c}{Pass@10} \\
 \hline 
Direct    & 48.1 & 55.1 & 64.7 & 75.0\\
Code CoT      & 53.9 & 56.4 & 63.5 & 68.6\\
Self-planning & 60.3 & 66.0 & 70.5 & 76.3\\ 
\hdashline
Ground-truth Planning  & 74.4 & 75.6 & 85.3 & 89.1\\ 
\bottomrule
\end{tabular}
}\label{sample}
\end{table}

\subsection{Performance on Different LLMs (RQ2)}

\textbf{Evaluation.} \, 
In this evaluation, we investigate the performance of self-planning approaches on different LLMs. We conduct experiments on the OpenAI language model family, including ada, cabbage, curie, cushman, and davinci. We use three 175B models—text-davinci-002, code-davinci-002, and text-davinci-003, which differ in training strategy and data. Furthermore, we apply the plan generated by code-davinci-002 during the planning phase to the implementation phase of other models, aiming to investigate the impact of planning for models with varying scales.
Since the input length limit of the small-size model is restrictive, we use the 4-shot setting for all prompting baselines in this experiment.

\begin{table*}[h!]
\caption{Performance of Self-planning approach across various base LLMs on HumanEval benchmarks.}
\centering
\resizebox{0.99\textwidth}{!}{
\begin{threeparttable}
\begin{tabular}{lcccccccccc}
\toprule
 \multirow{2}{*}{Approach} & \multicolumn{3}{c}{\cellcolor{gray!10} Direct} & \multicolumn{3}{c}{\cellcolor{gray!20} Self-planning} & \multicolumn{3}{c}{\cellcolor{gray!30} Planning} \\ \cmidrule(r){2-4} \cmidrule(r){5-7}  \cmidrule(r){8-10}
 & Pass@1 & CodeBLEU & AvgPassRatio & Pass@1 & CodeBLEU & AvgPassRatio & Pass@1 & CodeBLEU & AvgPassRatio\\ \hline
text-davinci-003 (175B)  & 55.1 & 31.5 & 72.2 & 65.4 & 29.6 & 80.1 & 65.4 & 30.2 & 80.9 \\
code-davinci-002 (175B)  & 48.1 & 24.0 & 63.2 & 59.0 & 29.3 & 77.8 & - & - & -\\
text-davinci-002 (175B)  & 48.1 & 24.4 & 63.1 & 50.0 & 28.8 & 69.4 & 57.1 & 30.4 & 76.3\\
code-cushman-001 (13B) & 34.0 & 20.8 & 53.2 & 30.1 & 23.1 & 50.5 & 44.9 & 26.7 & 67.1\\
\hdashline
text-curie-001 (6.7B) & 0.0 & 4.3 & 3.2 & 0.0 & 12.4 & 0.0 & 0.0 & 12.4 & 0.0\\
text-babbage-001 (1B) & 0.6 & 4.8 & 4.3 & 0.0 & 6.2 & 1.4 & 0.0 & 7.9 & 0.0\\
text-ada-001 (350M) & 0.0 & 3.9 & 0.9 & 0.0 & 7.4 & 0.2 & 0.0 & 7.8 & 0.1 \\
\bottomrule
\end{tabular}
 \begin{tablenotes}
        \item[1] Due to the maximum input length limitation, the evaluation of the self-planning is performed in the 4-shot setting.
        \item[2] We use code-davinci-002 for planning and the corresponding LLM for implementation.
 \end{tablenotes}
\end{threeparttable}
}\label{Scalling Law}
\end{table*}

\textbf{Results.} \, The experimental results are presented in Table \ref{Scalling Law}. When the model is small, the impact of self-planning is less pronounced, constrained by the model's inherent abilities. As the model size reaches 13B, the performance of LLMs in code generation begins to exhibit emerging ability, but self-planning ability remains relatively weak. At 175B, self-planning approach consistently outperforms the Direct approach across all models. For the same 175B model, code-davinci-002, fine-tuned on code, demonstrates a stronger self-planning ability than text-davinci-002. Furthermore, self-planning ability can be enhanced through reinforcement learning with human feedback (RLHF). It is evident that the self-planning ability of text-davinci-003 is significantly improved compared to text-davinci-002. Therefore, we posit that besides increasing model size, incorporating code training data and RLHF can also enhance the model's self-planning capabilities.

Subsequently, our experiments revealed that employing the plan generated by code-davinci-002 for models with lower abilities significantly improves their performance, particularly in the cases of code-cushman-001 and text-davinci-002. Text-ada-001, text-babbage-001, and text-curie-001 do not exhibit such performance as their inherent code generation ability is almost non-existent. An interesting observation is that when we utilize the plan generated by code-davinci-002 for the text-davinci-003 model, which is an upgraded version of the former, the resulting performance is approximately on par with text-davinci-003 for self-planning. This shows that text-davinci-003 does not improve the planning ability compared to code-davinci-002, what is improved is the code generation ability. 

In general, self-planning is an emergent ability that can only appear in large-enough language models; however, planning proves to be effective for most of the models.

\subsection{Variants of Self-planning (RQ3)}
\begin{figure}[h!]
\centering
\includegraphics[width=1\textwidth]{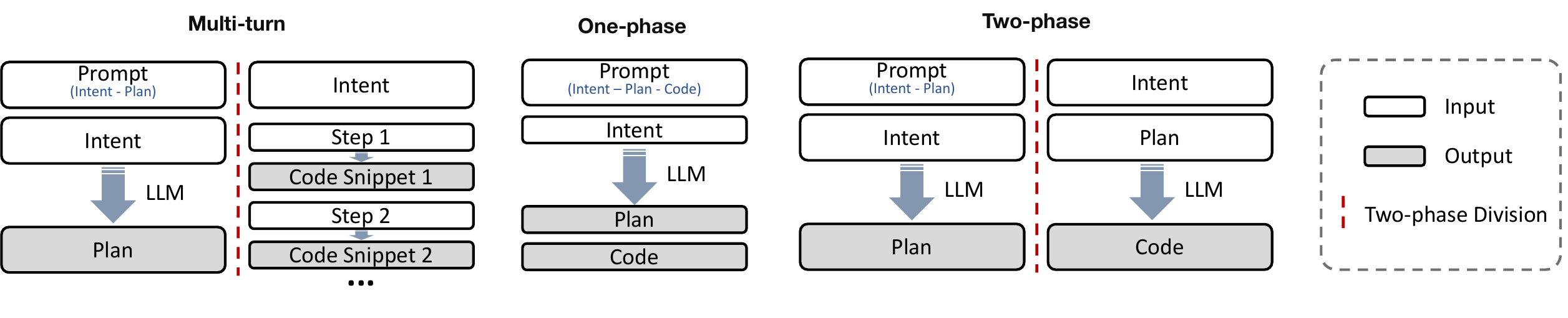}
\caption{Illustrations of the variants including two-phase, one-phase, and multi-turn.}
\label{variants_illustrations}
\end{figure}

\textbf{Evaluation.} \, 
We explore numerous variants in order to identify better choices for self-planning approach. First, we evaluate three planning and implementation schemes: multi-turn, one-phase, and two-phase. The illustrations of the variants including multi-turn, one-phase, and two-phase are shown in Fig.\ref{variants_illustrations}. The \textbf{Multi-turn} approach involves the iterative use of solution steps of plan to generate the corresponding code snippets that eventually compose the entire code, which is introduced by CodeGen \cite{nijkamp2022codegen}. In contrast, one-phase and two-phase schemes, both single-turn methods, employ all steps (i.e., the plan) to generate the entire code in a single iteration. However, while the \textbf{One-phase} approach simultaneously generates the plan and code, the \textbf{Two-phase} approach delineates these into separate phases.
Note that the one-phase approach requires labeling both the plan and the corresponding code in the prompt, as demonstrated in `Instance of Self-planning Prompt (One-phase)' in Appendix \ref{example_baseline}. Second, we evaluate the effects of self-planning with various example numbers (i.e., n-shot). Third, we explore six intermediate step configurations: Code CoT, Zero-shot CoT, narrative Code CoT, narrative plan, extremely concise plan, and Plan2CoT. \textbf{Zero-shot CoT} uses the instruction 'Let's think step by step' \cite{zeroshot_cot} to produce intermediate steps and code. To eliminate the effect of the generation method, we also establish the baseline \textbf{Zero-shot CoT (Two-phase)} for comparison, using the two-phase generation method consistent with our approach.
The \textbf{Narrative Plan} is an ablation of our proposed plan, i.e., it removes the form of a numbered list and is presented as narrative text. 
The \textbf{Narrative Code CoT} is also an ablation, which makes the Code CoT the same in form as the original CoT. The \textbf{Extremely Concise Plan} is an extremely concise version of our proposed plan. It is composed of only a few phrases or verbs (keep only the keywords as much as possible), and an example of it is displayed in Appendix \ref{example_baseline}. \textbf{Plan2CoT} means incorporate both the plan and the Code CoT during code generation, i.e., we first generate a plan and then generate a Code CoT, ultimately resulting in code. In this evaluation, in addition to evaluating the code generation quality using metrics Pass@1, CodeBLEU, and AvgPassRatio, we also measure the number of input/output tokens for the various variants. This approach helps understand the possible performance improvement in relation to possible cost increases.

\begin{table}[h!]
\setlength\tabcolsep{3pt}
\caption{Comparison of self-planning and its variants on HumanEval benchmark.}
\centering
\resizebox{0.99\textwidth}{!}{
\begin{tabular}{lcccc}
\toprule
Variant & Pass@1 & CodeBLEU & AvgPassRatio & Avg. Num. of I/O/I+O Tokens \\ \hline
\textbf{Schemes of Planning and Implementation}\\
Multi-turn & 28.2 & 18.1 & 47.5 & 2218.6 / 153.8 / 2372.4 \\
One-phase & 62.8 & 28.9 & 78.5 & 2354.5 / 105.9 / 2460.4 \\
\hdashline
\textbf{Number of Few-shot Example} \\
1-shot & 53.2 & 28.4 & 74.3 & 517.9 / 107.1 / 625.0 \\
2-shot & 54.8 & 26.7 & 74.4 & 700.4 / 111.1 / 811.5 \\
4-shot & 59.0 & 29.3 & 76.9 & 1047.8 / 113.7 / 1161.5 \\
\hdashline
\textbf{Configurations of Intermediate Step}\\
Zero-shot CoT & 18.0 & 12.6 & 37.6 & 276.8 / 78.8 / 355.6 \\
Zero-shot CoT (Two-phase) &50.6 & 23.8 & 70.9 & 505.5 / 381.8 / 887.3  \\
Narrative Code CoT & 50.6 & 29.0 & 72.2 & 2630.5 / 118.1 / 2748.6 \\
Narrative Plan & 55.8 & 27.7 & 73.6 & 1779.1 / 115.8 / 1894.9 \\
Extremely Concise Plan & 61.5 & 28.3 & 79.1 & 1669.8 / 84.7 / 1754.5 \\
Plan2CoT & 56.4 & 29.5 & 77.9 & 4547.9 / 185.4 / 4733.3 \\
\hdashline
Direct  & 48.1  & 24.0 & 63.2 & 130.6 / 41.5 / 172.1 \\
Code CoT & 53.9 & 30.4 & 75.6 & 2337.5 / 123.6 / 2461.1 \\
\textbf{Self-Planning (Two-phase, 8-shot)} & 60.3 & 28.6 & 80.8 & 1885.3 / 110.9 / 1996.2 \\ 
\bottomrule
\end{tabular}
}\label{variants}
\end{table}                          

\textbf{Results.} \, The results of the different variants on HumanEval benchmark are shown in Table \ref{variants}.

In the result of the group \textbf{Schemes of Planning and Implementation}, we find that the multi-turn usually fails to generate correct codes. This is attributed to the nature of LLMs. Since LLMs are trained on enormous concatenated texts and codes to predict the next token, LLMs may have truncation issues, i.e., they cannot precisely control the termination of their output. When using a plan to generate part of functions (usually several statements), it is difficult to define truncation rules. When implemented as a one-phase process, the self-planning approach has been shown to yield slightly improved performance compared to the two-phase way. However, this improvement is achieved at the cost of the increased complexity of crafting prompts. Specifically, the two-phase way only requires providing intent and plan examples in the prompt, whereas the one-phase way requires additional writing of the corresponding code examples. 

In the result of the group \textbf{Number of Few-shot Example}, we can observe that the performance of self-planning with n-shot improves as the value of n increases. However, it is crucial to consider the input length limit of LLMs (typically 2048 or 4096). As a result, it is not feasible to indefinitely increase the value of n without exceeding the input length limit. Considering the limitation of input length and the saturation of model performance growth, we generally recommend using either 8-shot or 4-shot for self-planning in LLMs. 

In the result of the group \textbf{Configurations of Intermediate Step}, the improvement of Code CoT over direct code generation is relatively small compared to the self-planning approach, as the challenge of generating an accurate and sufficiently detailed CoT is comparable to that of direct code generation. The performance with the Zero-shot CoT is worse than the few-shot prompt Code CoT we crafted. Thus CoT is not optimal for the code generation task, and the planning approach is more suitable. The degraded performance exhibited by narrative Code CoT and plan emphasizes the importance of clear, separated steps. For LLMs, consecutive and undifferentiated steps may lead to suboptimal understanding. Minor performance enhancement observed when self-planning approach employs an extremely concise plan reveals the powerful comprehension capabilities of LLMs, as well as the pivotal role that keywords play in the planning process. The performance of "Plan2CoT"  outperformed Code CoT approach, suggesting that planning prior to generating Code CoT can enhance the accuracy of Code CoT. However, it is slightly less effective than self-planning approach. We hypothesize that one layer of abstraction is sufficient for function-level code generation tasks in HumanEval. Relatively, excessive levels of abstraction may increase the probability of errors.

In terms of cost, self-planning has a lower token usage compared to other few-shot prompting approaches. It is particularly noteworthy that the total number of tokens used in 1-shot self-planning not only is lower than in Zero-shot CoT (Two-phase), but it also achieves higher performance. When using self-planning, users can make a trade-off between the number of tokens and performance as needed, opting to use different numbers of shots.

Overall, all variants except one-stage and extremely concise plan underperform the self-planning approach in terms of performance. However, one-stage approach necessitates the provision of code within the prompt, presenting a barrier to entry for individuals lacking programming expertise, while extremely concise plan approach does not align with human writing conventions. Taking all these factors into account, our proposed planning method performs as the optimal choice among the variants we explored.

\subsection{Performance on Multilingual Code Generation (RQ4)}

\textbf{Evaluation.} \, 
We evaluate the generality of our self-planning approach on HumanEval-X benchmarks across multiple PLs, i.e. Python, Java, Javascript, and Go. Rather than customizing plans for each specific PL, we utilize the same intent-plan pair example as Python across all PLs.

\begin{table}[h!]
\caption{Comparison of self-planning and other approaches on multilingual Datasets.}
\centering
\setlength\tabcolsep{2pt}
\small{
\begin{tabular}{lcccc}
\toprule
\multirow{2}{*}{Approach} & 
\multicolumn{2}{c}{\cellcolor{gray!10} Python} & \multicolumn{2}{c}{\cellcolor{gray!20} Java}\\
\cmidrule(r){2-3} \cmidrule(r){4-5} 
\multicolumn{1}{l}{} &
\multicolumn{1}{c}{Pass@1} &
\multicolumn{1}{c}{CodeBLEU} & \multicolumn{1}{c}{Pass@1} &
\multicolumn{1}{c}{CodeBLEU} \\
\hline 
Direct    & 48.1  & 24.0 & 50.6 & 38.0\\
Code CoT      & $53.9 \ \ (\uparrow 12.1\%)$  & 30.4 & $56.4 \ \ (\uparrow 11.5\%)$ & 39.0\\
Self-planning & $\textbf{60.3}\ \ (\textcolor{red}{\uparrow 25.4\%})$  & 28.6 & $\textbf{61.5}\ \ (\textcolor{red}{\uparrow 21.5\%})$ & 39.0\\ 
\hdashline
Ground-truth Planning  & $74.4 \ \ (\textcolor{teal}{\uparrow 54.7\%})$ & 41.0 & $66.7 \ \ (\textcolor{teal}{\uparrow 31.8\%})$ & 45.8\\ 

\hline  
\multicolumn{1}{l}{} & 
\multicolumn{2}{c}{\cellcolor{gray!10} Javascript} & \multicolumn{2}{c}{\cellcolor{gray!20} Go} \\
Direct    & 53.2 & 26.7  & 42.9 & 22.2\\
Code CoT     & $52.6  \ \ (\uparrow -1.1\%)$ & 27.0 & $48.1 \ \ (\uparrow 12.1\%)$ & 27.1\\
Self-planning & $\textbf{55.8}\ \ (\textcolor{red}{\uparrow 4.9\%})$ & 25.6 & $\textbf{53.0}\ \ (\textcolor{red}{\uparrow 23.5\%})$ & 26.5 \\ 
\hdashline
Ground-truth Planning  & $60.3 \ \ (\textcolor{teal}{\uparrow 13.3\%})$ & 29.6 & $58.3 \ \ (\textcolor{teal}{\uparrow 35.9\%})$ & 32.0 \\ 

\bottomrule
\end{tabular}
}\label{multilingual_results}
\end{table}

\textbf{Results.} \, As demonstrated in Table \ref{multilingual_results}, our self-planning approach exhibits positive results across all PLs when compared to Direct and Code CoT. It is evident that our method yields the most significant improvement for Python. This may be due to the fact that we tend to solve in Python when writing plans, introducing some of the features that are common to Python, such as dict, list, etc. As a result, the improvements are more pronounced for PLs that have Python-like features, such as Go and Java. We believe that if plans are customized for other PLs, their effectiveness would be further enhanced. Moreover, if a plan is created independent of a specific PL, its generalization across different languages would be improved.

\subsection{Effect of Problem Complexity on Self-planning}

\textbf{Evaluation.} \, 
In this section, we focus on evaluating the impact of problem complexity on the performance of self-planning approach. We split the questions in HumanEval according to their difficulty. Specifically, we followed the LLMs evaluation methodology \cite{DBLP:conf/sbes/0001CMLSG23} by providing the intent and code examples of each problem for GPT-4 and using a specific instruction $I$ to score the difficulty of the problem. Thereafter, we perform a full manual review of the scoring results. Finally, we categorize the HumanEval dataset into three difficulty levels based on a segmentation function $f$.
\[
I = 
\begin{minipage}{0.8\linewidth}
\textit{``Please rate the level of difficulty of this problem with an integer between 0 and 10, where 0 is the easiest, and 10 is the hardest. Return the response in a JSON format, with a variable ’score’ containing your score, and another variable ’explanation’ with the explanation for this score. Your explanation must have at least 20 words.''}
\end{minipage}
\]
\[
\text{difficulty} = f(score) =
\begin{cases} 
  1 & \text{if } 0 <  score \leq 2 \\
  2 & \text{if }  score = 3   \\
  3 & \text{if }  score \geq 4 \\
\end{cases}
\]
The reason we choose to use this segmentation function is that by splitting the questions in this way, we are able to make the number of questions contained in each difficulty level more balanced. The categorization results are: difficulty level 1 contains 67 questions, difficulty level 2 contains 47 questions, and difficulty level 3 contains 43 questions. After splitting the dataset in this manner, we evaluated the Pass@1 metric for three approaches: Direct, Code CoT, and Self-planning.

\begin{table}[h!]
\caption{Pass@1 of self-planning on problems of varying complexity.}
\centering
\small{
\begin{tabular}{lccc}
\toprule
\multicolumn{1}{l}{\multirow{1}{*}{Approach}} & 
  \multicolumn{1}{c}{Complexity 1} &
  \multicolumn{1}{c}{Complexity 2} &
  \multicolumn{1}{c}{Complexity 3} \\
 \hline 
Direct    & 65.2 & 36.2 & 31.1 \\
Code CoT      & $68.1 \ \ (\uparrow 4.4\%)$ & $44.7 \ \ (\uparrow 23.4\%)$ & $37.3 \ \ (\uparrow 19.9\%)$\\
\hdashline
Self-planning & $71.2 \ \ (\textcolor{red}{\uparrow 9.3\%})$ & $55.3 \ \ (\textcolor{red}{\uparrow 52.7\%})$ & $50.2 \ \ (\textcolor{red}{\uparrow 61.5\%})$\\ 
\bottomrule
\end{tabular}
}\label{complexity}
\end{table}

\textbf{Results.} \, 
The results of the evaluation are shown in Table \ref{complexity}. The self-planning approach exhibits higher performance than all baseline approaches, Direct and Code CoT, when dealing with problems of varying difficulty. It is worth noting that the performance improvement of the self-planning approach is particularly significant when dealing with higher difficulty problems (Complexity 2, Complexity 3), exceeding the improvement on lower difficulty problems (Complexity 1). This observation underscores that the Self-planning approach excels in handling complex problems, while simultaneously maintaining efficient performance for simpler tasks.

\section{Human Evaluation}
In this section, we conduct a human evaluation to assess the quality of the self-planning and baseline approaches. This evaluation is designed to reflect the practical experience of human developers using these code generation approaches. The results of this evaluation will provide valuable insights into the usability and practicality of self-planning code generation.

\textbf{Evaluation.} \, We first establish a set of criteria to assess the generated code, as outlined below.
\begin{itemize}
    \item Correctness: High-quality code should be correct and produce the expected output or behavior. This means that the code should meet the requirements, and its functionality should be accurate and precise.
    \item Readability: High-quality code should be easy to read and understand by developers, facilitating future maintenance. This can be achieved through clear naming conventions, consistent indentation and formatting, and using comments to explain complex or unclear sections.
    \item Robustness: High-quality code should be robust and handle unexpected situations or edge cases gracefully.
\end{itemize}
Second, we sample 50 tasks from the HumanEval benchmark, and each task contains five codes: ground-truth code, direct-generated code, self-planning-generated code, Code CoT-generated code, and ground-truth planning-generated code. We asked developers to score each code on five aspects from the criteria. The scores are integers ranging from 0 to 4, where 0 is the worst and 4 is the best. Note that we show developers five codes for one task at a time, making it easy for developers to compare the five codes and score the gaps.

Finally, we assemble a team of evaluators, including 10 developers with 2-5 years of Python programming experience, and divide them into two evaluation groups (Group A and Group B). The evaluation is conducted in the form of an anonymous questionnaire, which is displayed in the Appendix \ref{questionnaire}. Each evaluation team is required to evaluate all tasks, and each evaluator is randomly assigned 10 tasks (questionnaires), where the codes generated by the different methods corresponding to each task are randomly ordered.

\begin{figure}[h!]
\centering
\includegraphics[width=0.8\textwidth]{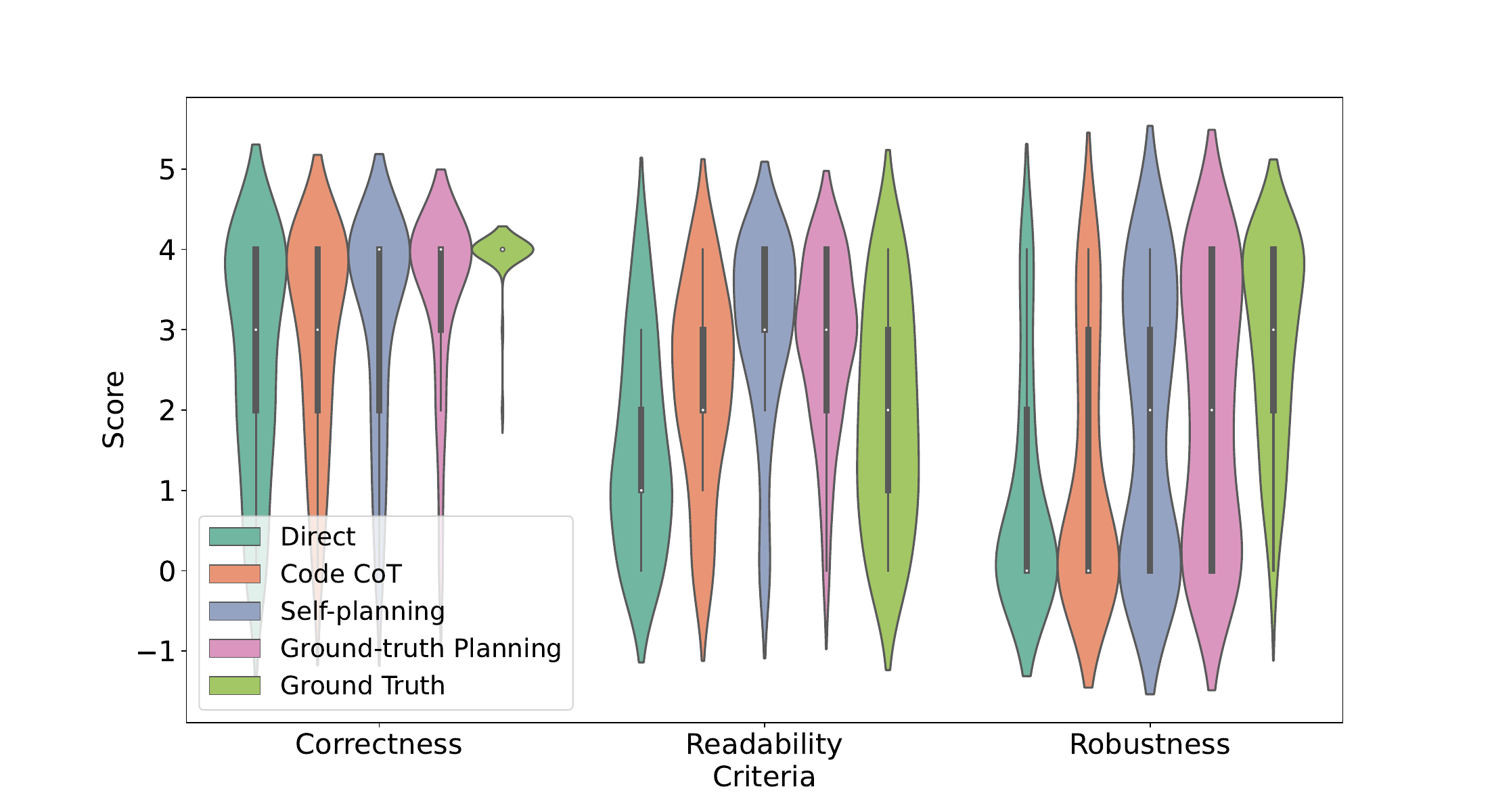}
\caption{Violin plot for human evaluation. The violin plot is a combination of a box line plot, which shows the location of the quartiles (the upper edge of the box represents the upper quartile, the middle point is the median, and the lower edge of the box is the lower quartile), and a kernel density plot, which shows the density at any location.}
\label{violin}
\end{figure}

\textbf{Results.} \, 
The evaluation results of the two groups are summarized in Fig. \ref{violin}.
Correctness scoring closely aligns with the Pass@1 results, while also considering partial correctness. The self-planning approach outperforms both Direct and Code CoT but falls short of Ground-truth planning and Ground-truth.

In terms of readability, the self-planning approach excels as the most readable, distinguished by a blend of accuracy and a coherent outline of its planned steps. The Ground-truth planning approach's readability closely follows that of the self-planning approach. Conversely, the Code CoT exhibits subpar readability. Its solution steps provide excessive detail, potentially hindering the identification of crucial elements within the code and becoming outdated if code modifications occur. This can adversely affect code maintainability if the solution steps do not accurately represent the current state. The readability of both Direct and Ground-truth is deemed insufficient.

\begin{figure*}[t!]
\centering
\includegraphics[width=1\textwidth]{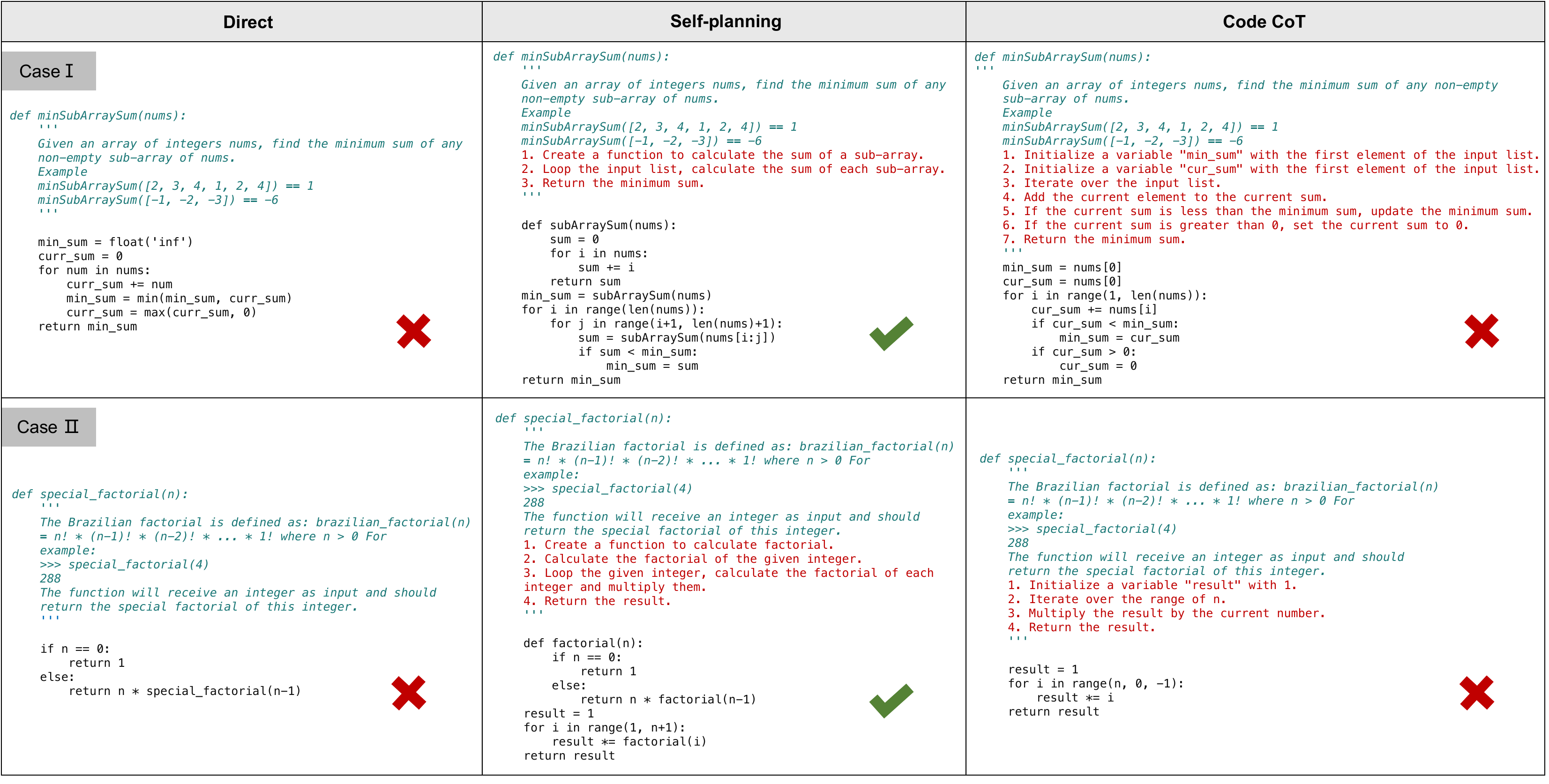}
\caption{Two real cases from HumanEval with self-planning, Code CoT (with self-planning format), and direct code generation. The input, generated plan, and code are highlighted in green, red, and black, respectively.}
\label{cases}
\end{figure*}

We find that the incorrect code usually receives a score of 0 for robustness item. Consequently, the robustness of violin plots displays a broader pattern at the lower end. Regarding robustness, the self-planning approach surpasses Code CoT and performs comparably to Ground-truth planning, since the self-planning approach can thoroughly consider some edge cases and determines the legality of inputs, as evidenced by the qualitative examples \ref{cases}. 

In conclusion, through human evaluation, our self-planning approach exhibits the best readability among all approaches, and its correctness and robustness performance is on par with the ground-truth planning approach.

\textbf{Qualitative Examples.} \, 
To further understand the effectiveness of our approach, we examine cases where self-planning approach has contrasting performances to Direct approach. As depicted in Fig. \ref{cases}, we demonstrate the performance of both direct, self-planning, and Code CoT code generation through two cases. In these cases, the direct and Code CoT code generation approach only addresses a limited aspect of intent, which often results in incorrect code generation. In contrast, the self-planning code generation approach first converts the intent into plan, and then systematically resolves each solution step of plan. This approach effectively minimizes the risk of overlooking critical elements. 

In case \uppercase\expandafter{\romannumeral1}, the task of LLMs is “\textit{Given an array of integers nums, find the minimum sum of any non-empty sub-array of nums}”. The code generated directly by LLM only considers a subset of the sub-arrays, whereas our approach ensures that none of them are overlooked.
In case \uppercase\expandafter{\romannumeral2}, the task of LLMs is “\textit{Receive an integer as input and return the special factorial of this integer.}” The direct code generation simply implements the standard factorial in a recursive manner, neglecting the definition of the special factorial. In contrast, our approach implements the standard factorial through the use of sub-function and subsequently uses the sub-function to construct the special factorial.
We can find that the semantics of the code generated by Code CoT and direct code generation are almost the same, only the expression form is different. This may confirm the point that the difficulty of generating Code CoT from intent and generating code is comparable.

Overall, our self-planning approach offers a more thorough and nuanced way for addressing complex tasks assigned to LLMs, in contrast to direct and Code CoT code generation, which provides a more straightforward and limited solution.

\section{Threats to Validity}
There are two primary threats to the validity of our study.

\begin{itemize}
    \item Considering the inherent sensitivity of LLMs to prompts, the primary threat comes from crafting prompts, as the example selection and plan writing in this operation can affect the degree of improvement achievable by our approach. This issue necessitates fundamental improvements to the LLMs for resolution \cite{NookalaVMK23, promptbench}. In our current approach, the random sampling of examples and adherence to plan-writing principles ensure a considerable level of improvement. However, there is potential for optimization. Several research efforts have explored how automated techniques for selecting quality examples \cite{RetrievePrompts} and generating prompts \cite{AutomaticChain,PromptEngineers} can be used to maximize the performance of the prompting approach. These results can be introduced into our approach to further improve the performance.

    \item The second major threat to our study pertains to the generalizability of experimental results. To address this threat, we assessed self-planning on seven public benchmark datasets, in line with previous work \cite{codex, codegeex, CodeScore}. These datasets span four mainstream programming languages: Python, Java, Javascript, and Go. To validate the quality of the generated code, we employed the widely accepted metric, Pass@k, which leverages test cases to gauge the functional correctness of code. Additionally, we utilized the unbiased version of Pass@k \cite{codex} to diminish evaluation errors that arise from sampling.
\end{itemize}

\section{Related Work}
\subsection{Code Generation}
Traditional code generation approaches are based on supervised learning, which initially treats code as equivalent to natural language \cite{LingBGHKWS16,JiaL16,WeiBolin} and then gradually incorporates more code-specific features, such as abstract syntax tree \cite{RabinovichSK17,YinN17,SunZMXLZ19,SunZXSMZ20,TranX}, API calls \cite{RaychevVY14,GuZZK16,GuZZ017}. Furthermore,  Mukherjee et al.\cite{MukherjeeWCRCJ21} present a generative modeling approach for source code that uses a static analysis tool. Dong et al. \cite{CODEP} devise a PDA-based methodology to guarantee grammatical correctness for code generation. 

With the rise of pre-training, CodeT5 \cite{CodeT5}, UniXcoder \cite{UniXcoder} applied pre-trained models to code generation task. The introduction of subsequent models like Codex \cite{codex}, InCoder \cite{incoder}, CodeGen \cite{nijkamp2022codegen}, AlphaCode \cite{alphacode}, and CodeGeeX \cite{codegeex}, continues to push the direction. A noteworthy trend among these models is the rapid increase in the number of parameters in the pre-trained models, which leads to a tremendous improvement in the performance of code generation \cite{CDD}. 
This has sparked a variety of studies, with the focus on utilizing large language models as the backbone, enhancing their code generation performance through various techniques \cite{codet, CoderReviewer, planningdecoding, debug}, and achieving promising results. These approaches can be summarized as post-processing strategies, i.e., reranking and modifying the code after the model generates it. Our approach and post-processing approaches are orthogonal and can be used concurrently.

\subsection{Prompting Techniques}
\label{Related work}
Few-shot prompting \cite{LiuYFJHN23} is a technique that emerged as the number of model parameters exploded. Instead of fine-tuning a separate language model checkpoint for each new task, few-shot prompting can be utilized by simply providing the model with a limited number of input-output examples that illustrate the task. A few-shot prompt technique known as Chain of thought (CoT) \cite{wei2022chain} achieves a significant improvement that transcends the scaling laws by generating intermediate reasoning steps before the answer to address language reasoning tasks. \textbf{CoT in its original paper focused on solving a range of reasoning tasks such as mathematical reasoning, symbolic reasoning, and commonsense reasoning, and did not implement CoT for code-generation tasks.} Inspired by CoT, a series of prompting works has been proposed. Least-to-most prompting \cite{leastmost} reduces a complex problem into a series of sub-problems and then solves the sub-problems in order, adding the answer to the previous sub-problems to the prompt each time solving begins. PAL \cite{pal} and PoT \cite{POT} are proposed to generate code as the intermediate reasoning steps, delegating solving to the compiler, thus improving solution accuracy. Nonetheless, the aforementioned approaches are adept at addressing relatively simple mathematical \cite{LewkowyczADDMRS22,WuJLRSJS22}, commonsense \cite{SanhWRBSACSRDBX22,MadaanZ0YN22},  and symbolic reasoning \cite{YaoZYDSN023} problems characterized by limited problem spaces and established solution patterns. Consequently, their applicability to code generation remains restricted.

\subsection{Self-improvement of LLMs}
The use of LLMs to enhance the performance of LLMs themselves has become a current research hotspot. Huang et al. \cite{Huang} demonstrate that an LLM can enhance its performance on reasoning datasets by training on the data it generated itself. Self-Instruct \cite{Self-Instruct} improves the instruction-following capabilities of LLMs by bootstrapping off their own generations. Self-Refine \cite{self-refine} uses the LLMs to provide feedback for its own output and refine itself. Self-Evaluation \cite{self-evaluation} introduces a stepwise self-evaluation mechanism to guide and calibrate the reasoning process of LLMs. Self-validation \cite{Self-Verification} improves few-shot clinical information extraction by utilizing the LLM to provide provenance for its own extractions and checking its own output. Self-Criticism \cite{Self-Criticism} achieves alignment with being helpful, honest, and harmless by utilizing LLMs to judge themselves. Beyond these, Promptbreeder \cite{Promptbreeder} utilizes a self-referential self-improvement mechanism with LLMs that evolves and adapts prompts for a specific domain.

\subsection{Planning with LLMs}
The practice of utilizing LLMs for planning in various applications is increasingly garnering attention, particularly for their potential to comprehend complex problems and optimize decision-making processes.
LLM-Planner \cite{LLM-Planner} harnesses the power of large language models to do few-shot planning for embodied agents. 
HuggingGPT \cite{Hugginggpt} integrates LLMs to plan the invocation of various AI models from the machine learning community (e.g., Hugging Face) for handling a wide range of complex AI tasks spanning different paradigms and domains.
DEPS \cite{DEPS} proposes an interactive planning approach based on LLMs capable of robustly completing more than 70 Minecraft tasks.
ProgPrompt \cite{ProgPrompt} leverages LLMs to generate situated robot task plans.

\section{Discussion and Future Work}
When discussing the limitations of self-planning code generation, a major limitation may be the manual crafting of prompts. However, we should also be aware that in previous approaches, a huge number of examples may be needed in order to train a model to understand planning. This makes data efficiency an important issue. However, we propose a self-planning approach that directly teaches LLMs to understand planning with only a few examples and can be crafted by people without programming knowledge. This improvement in data efficiency and low barrier can make the self-planning code generation approach easier to apply in practice.

Additionally, our approach employs an approximate sequential executed list to represent plans, which is similar to the functional points delineated in requirements documents. Considering the powerful capabilities of LLMs, some rudimentary loops and branch structures in the plan may not be necessary. For instance, in Fig. \ref{self-plan} and \ref{cases}, the sub-functions encompass loops and branch structures. However, we merely need to specify the function of these sub-functions without describing the internal loop and branching structure. Some intricate plans can be simplified through iterative planning until they no longer pose challenges for LLM.

Finally, this paper attempts to reduce the difficulty of code generation by planning for human intent, which is consistent with the methodology of dealing with problem complexity in requirements engineering, i.e. abstraction and decomposition \cite{macaulay2012requirements}. The current LLMs are capable of generating code that addresses simple human requirements, however, it is still a long way from producing a fully functional piece of software. The requirements of software development are significantly more complex and intricate. It may be worthwhile to explore beyond code writing to the realm of requirements analysis, incorporating the methodology of requirements engineering with LLMs.

\section{Conclusion}
In this paper, we have explored plan-aided code generation and proposed a simple but effective approach to perform self-planning and generate code with LLMs. Self-planning code generation outperforms direct generation with LLMs on multiple code generation datasets by a large margin. Moreover, self-planning approach leads to enhancements in the correctness, readability, and robustness of the generated code, as evidenced by human evaluation. Empirical evidence indicates that although self-planning is an emergent ability, incorporating planning strategies can yield advantages for most models.

\begin{acks}
This research is supported by the National Key R\&D Program under Grant No. 2023YFB4503801, the National Natural Science Foundation of China under Grant No. 62072007, 62192733, 61832009, 62192730, and the Key Program of Hubei under Grant JD2023008.
\end{acks}

\newpage
\bibliographystyle{ACM-Reference-Format}
\bibliography{ref}

\newpage
\appendix

\section{Vanilla COT}\label{VanillaCOT}
\begin{figure}[h!]
\centering
\includegraphics[width=0.83\textwidth]{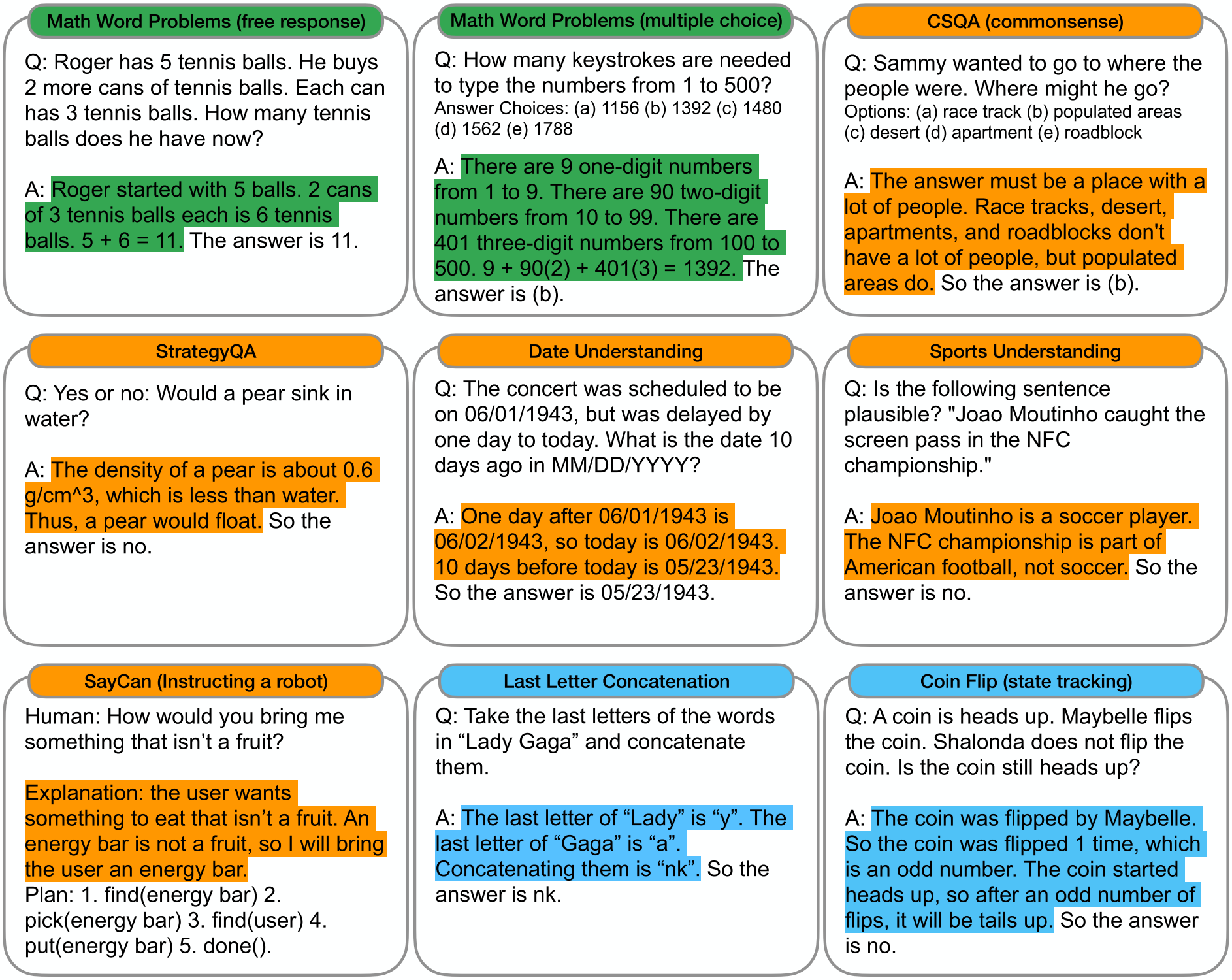}
\caption{Examples of <input, CoT, output> triples for arithmetic, commonsense, and symbolic reasoning benchmarks. CoT are highlighted. The examples are from the original paper of COT \cite{wei2022chain}.}
\label{vCoT}
\end{figure}

Vanilla CoT focus on solving natural language reasoning tasks such as mathematical reasoning, symbolic reasoning, and commonsense reasoning, and does not apply to code-generation tasks. Examples of Vanilla CoT is shown in Fig. \ref{vCoT}.

\section{Code COT}
\begin{figure}[h!]
\centering
\includegraphics[width=0.7\textwidth]{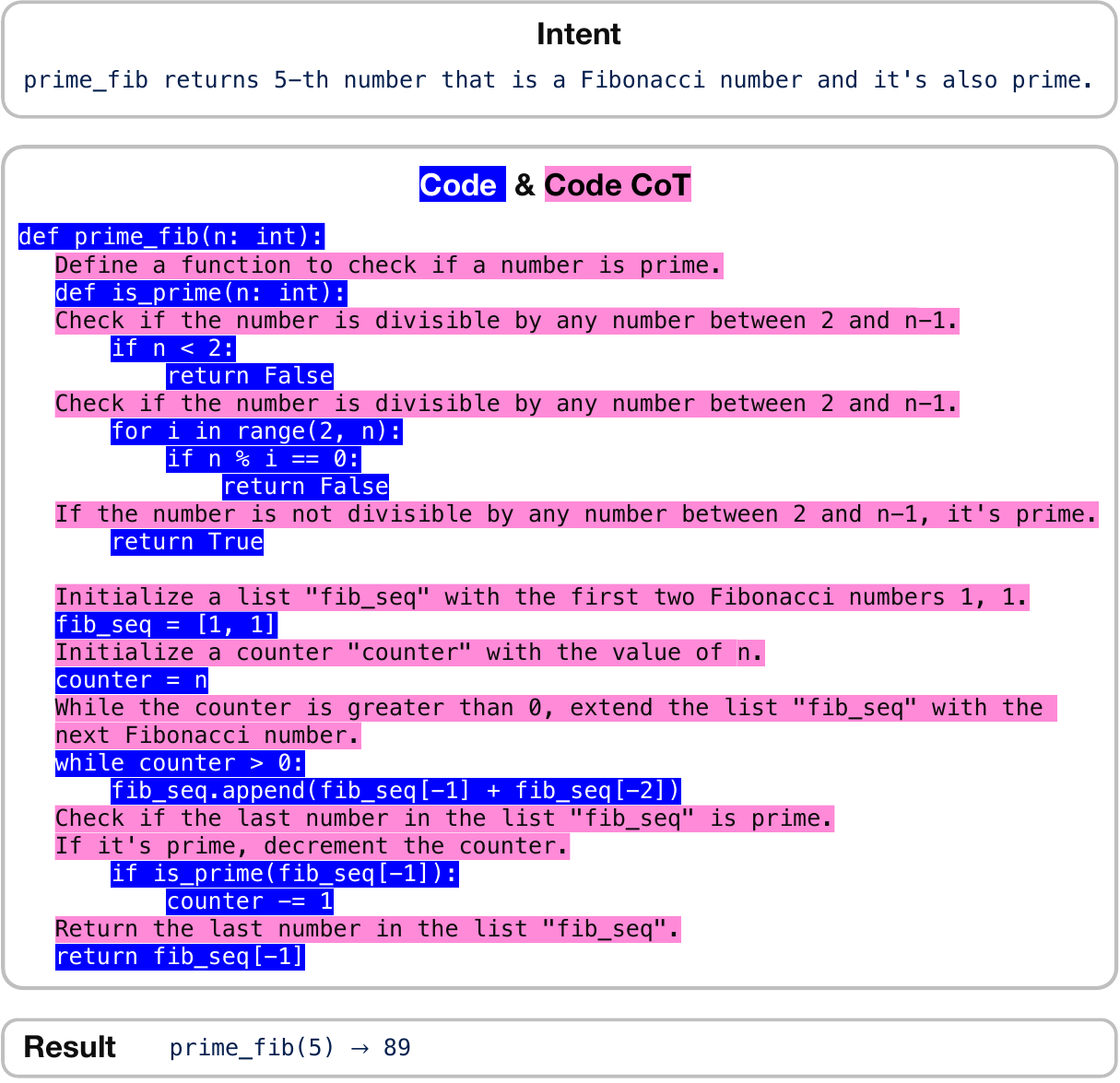}
\caption{An example of Code CoT compared to code, where we alternate the display between Code and Code CoT to make it easier to understand their similarities. However, in practice, we let LLM generate a complete Code CoT directly from intents, and then generate code from Code CoT. \textbf{Note that CoT has not been applied to code generation before. Here we implement Code CoT as an earliest adaptation of CoT for code generation.}}
\label{planCoT}
\end{figure}

Following CoT, we implement it for code generation. An example of applying CoT to code generation is shown in Fig. \ref{planCoT}.

\section{Details of Metrics}
\label{Details of Metrics}

\textbf{Pass@k.} \ We use the unbiased version \cite{alphacode} of Pass@k, where $n>=k$ samples are generated for each problem, count the number of correct samples $c<=n$ which pass test cases, and calculate the following estimator,
\begin{equation}
    \operatorname{Pass@k} = \mathop{\mathbb{E}}\limits_{\operatorname{Problems}}\begin{bmatrix}1-\tiny{\frac{\begin{pmatrix}n-c\\k\end{pmatrix}}{\begin{pmatrix}n\\k\end{pmatrix}}}\end{bmatrix}.
\end{equation}

\textbf{AvgPassRatio.} \ The average proportion of test cases \cite{apps} that generated codes $\mathbf{g}_p's$ pass: 
\begin{equation}
    \label{AvgPassRatio}
    \frac{1}{|P|} \sum_{p\in P} \frac{1}{|C_{p}|} \sum_{c\in C_{p}} \mathbb{I}\left\{\operatorname{Eval}\left(\mathbf{g}_p, \mathcal{I}_{p,c} \right)=\mathcal{O}_{p,c}\right\},
\end{equation}
where $|\cdot|$ indicates the cardinality of a set, $\mathbb{I}(\cdot)$ is an indicator function, which outputs 1 if the condition is true and 0 otherwise, and $\operatorname{Eval}\left(\mathbf{g}_p, \mathcal{I}_{p,c} \right)$ represents an evaluation function that obtains outputs of code $\mathbf{g}_p$ by way of executing it with $\mathcal{I}_{p,c}$ as input. 

\textbf{CodeBLEU.} \ CodeBLEU \cite{codebleu} is a variant of BLEU that injects code features for code evaluation. CodeBLEU considers abstract syntax tree and dataflow matching in addition to n-gram co-currency (BLEU),
\begin{align*}
    \text { CodeBLEU }& = \alpha \cdot \operatorname{BLEU} + \beta \cdot \operatorname{BLEU}_{w e i g h t}\\  & + \delta \cdot \text{Match}_{ast} + \zeta \cdot \text{Match}_{df}.
\end{align*}

\section{Experimental Verification of Subproblem Decomposition in Self-planning}
{
We conduct additional experiments to further verify the decomposition of subproblems by our approach. We count the number of generated sub-functions (denotes a successfully decomposed sub-problem) for Direct, Code CoT (i.e., the CoT implemented for code generation in this paper), and Self-planning on the HumanEval dataset. The results show that Direct, Code CoT, and Self-planning generate 1, 9, and 37 sub-functions, respectively. Moreover, we pick the multi-objective problems (e.g., ``prime\_fib returns n-th number that is a Fibonacci number and it's also prime.'') that decomposition is skilled at solving. The correct number of multi-objective problems (total of 58) using Direct, Code CoT, and Self-planning are 11, 18, and 40, respectively. These experiments indicate that self-planning helps break down the problem into sub-problems, thereby facilitating code generation.}

\section{Self-planning Prompt for HumanEval Benchmarks} \label{self_planning_prompt_for_humanEval}

\begin{mybox}

\noindent def encrypt(s):\\ 
    \indent\qquad '''\\
    \indent\qquad Create a function encrypt that takes a string as an argument and returns a string encrypted with the alphabet being rotated. The alphabet should be rotated in a manner such that the letters shift down by two multiplied to two places.\\
    \indent\qquad For example:\\
    \indent\qquad encrypt('hi') returns 'lm'\\
    \indent\qquad encrypt('asdfghjkl') returns 'ewhjklnop'\\
    \indent\qquad encrypt('gf') returns 'kj'\\
    \indent\qquad encrypt('et') returns 'ix'\\
    \indent\qquad 1. Create a alphabet, bias two places multiplied by two.\\
    \indent\qquad 2. Loop the input, find the latter bias letter in alphabet.\\
    \indent\qquad 3. Return result.\\
    \indent\qquad '''
    
\noindent \raisebox{1ex}{\rule{\linewidth}{0.2mm}}

\noindent def check\_if\_last\_char\_is\_a\_letter(txt):\\
    \indent\qquad '''\\
    \indent\qquad Create a function that returns True if the last character of a given string is an alphabetical character and is not a part of a word, and False otherwise. Note: "word" is a group of characters separated by space.\\
    \indent\qquad Examples:\\
    \indent\qquad check\_if\_last\_char\_is\_a\_letter("apple pie") $\rightarrow$ False\\
    \indent\qquad check\_if\_last\_char\_is\_a\_letter("apple pi e") $\rightarrow$ True\\
    \indent\qquad check\_if\_last\_char\_is\_a\_letter("apple pi e ") $\rightarrow$ False\\
    \indent\qquad check\_if\_last\_char\_is\_a\_letter("") $\rightarrow$ False \\
    \indent\qquad 1. If the string is empty, return False.\\
    \indent\qquad 2. If the string does not end with a alphabetical character, return False.\\
    \indent\qquad 3. Split the given string into a list of words.\\
    \indent\qquad 4. Check if the length of the last word is equal to 1.\\
    \indent\qquad '''
  
\noindent \raisebox{1ex}{\rule{\linewidth}{0.2mm}}

\noindent def file\_name\_check(file\_name):\\
    \indent\qquad '''\\
    \indent\qquad Create a function which takes a string representing a file's name, and returns 'Yes' if the the file's name is valid, and returns 'No' otherwise. A file's name is considered to be valid if and only if all the following conditions are met: - There should not be more than three digits ('0'-'9') in the file's name. - The file's name contains exactly one dot '.' - The substring before the dot should not be empty, and it starts with a letter from the latin alphapet ('a'-'z' and 'A'-'Z'). - The substring after the dot should be one of these: ['txt', 'exe', 'dll']\\
    \indent\qquad Examples:\\
    \indent\qquad file\_name\_check("example.txt")  =$>$ 'Yes'\\
    \indent\qquad file\_name\_check("1example.dll")  =$>$ 'No' (the name should start with a latin alphapet letter)\\
    \indent\qquad 1. Check if the file name is valid according to the conditions.\\
    \indent\qquad 2. Return "Yes" if valid, otherwise return "NO".\\
    \indent\qquad '''
  
\noindent \raisebox{1ex}{\rule{\linewidth}{0.2mm}}

\noindent def fruit\_distribution(s,n): \\
    \indent\qquad ''' \\
    \indent\qquad In this task, you will be given a string that represents a number of apples and oranges that are distributed in a basket of fruit this basket contains apples, oranges, and mango fruits. Given the string that represents the total number of the oranges and apples and an integer that represent the total number of the fruits in the basket return the number of the mango fruits in the basket. \\
    \indent\qquad for examble: \\
    \indent\qquad fruit\_distribution("5 apples and 6 oranges", 19) ->19 - 5 - 6 = 8\\ 
    \indent\qquad fruit\_distribution("0 apples and 1 oranges",3) -> 3 - 0 - 1 = 2 \\
    \indent\qquad fruit\_distribution("2 apples and 3 oranges", 100) -> 100 - 2 - 3 = 95 \\
    \indent\qquad fruit\_distribution("100 apples and 1 oranges",120) -> 120 - 100 - 1 = 19\\
    \indent\qquad 1. Extract the numbers of oranges and apples from given string.\\
    \indent\qquad 2. Calculate the sum of oranges and apples.\\
    \indent\qquad 3. Deduct the sum from the total number of fruits.\\
    \indent\qquad 4. Return the number of mangoes.\\
    \indent\qquad '''

\noindent \raisebox{1ex}{\rule{\linewidth}{0.2mm}}

\noindent def prime\_fib(n: int):  \\
    \indent\qquad '''  \\
    \indent\qquad prime\_fib returns n-th number that is a Fibonacci number and it's also prime. \\
    \indent\qquad Examples: \\
    \indent\qquad $>>>$ prime\_fib(1) 2  \\
    \indent\qquad $>>>$ prime\_fib(2) 3  \\
    \indent\qquad $>>>$ prime\_fib(3) 5  \\
    \indent\qquad $>>>$ prime\_fib(4) 13  \\
    \indent\qquad $>>>$ prime\_fib(5) 89 \\
    \indent\qquad 1. Create a function to check if a number is prime. \\
    \indent\qquad 2. Generate a Fibonacci sequence. \\
    \indent\qquad 3. Check if each number in the Fibonacci sequence is prime, decrement the counter. \\
    \indent\qquad 4. If the counter is 0, return the Fibonacci number. \\
    \indent\qquad '''
    
\noindent \raisebox{1ex}{\rule{\linewidth}{0.2mm}}

\noindent def compare\_one(a, b): \\
    \indent\qquad '' \\
    \indent\qquad Create a function that takes integers, floats, or strings representing real numbers, and returns the larger variable in its given variable type. Return None if the values are equal. Note: If a real number is represented as a string, the floating point might be . or , \\
    \indent\qquad Examples: \\
    \indent\qquad compare\_one(1, 2.5) $\rightarrow$ 2.5 \\
    \indent\qquad compare\_one(1, "2,3") $\rightarrow$ "2,3" \\
    \indent\qquad compare\_one("5,1", "6") $\rightarrow$ "6" \\
    \indent\qquad compare\_one("1", 1) $\rightarrow$ None \\
    \indent\qquad 1. Store the original inputs. \\
    \indent\qquad 2. Check if inputs are strings and convert to floats. \\
    \indent\qquad 3. Compare the two inputs and return the larger one in its original data type. \\
    \indent\qquad '''
    
\noindent \raisebox{1ex}{\rule{\linewidth}{0.2mm}}

\noindent def sort\_even(l: list): \\
    \indent\qquad ''' \\
    \indent\qquad This function takes a list l and returns a list l' such that l' is identical to l in the odd indicies, while its values at the even indicies are equal to the values of the even indicies of l, but sorted. \\
    \indent\qquad Examples: \\
    \indent\qquad $>>>$ sort\_even([1, 2, 3]) \\
    \indent\qquad [1, 2, 3] \\
    \indent\qquad $>>>$ sort\_even([5, 6, 3, 4]) \\
    \indent\qquad [3, 6, 5, 4] \\
    \indent\qquad 1. Create a list of all the even indices of the given list. \\
    \indent\qquad 2. Sort the list of even indices. \\
    \indent\qquad 3. Return a new list that is identical to the original list in the odd indicies, and equal to the sorted even indices in the even indicies. \\
    \indent\qquad '''
    
\noindent \raisebox{1ex}{\rule{\linewidth}{0.2mm}}

\noindent def search(lst): \\
    \indent\qquad ''' \\
    \indent\qquad You are given a non-empty list of positive integers. Return the greatest integer that is greater than zero, and has a frequency greater than or equal to the value of the integer itself. The frequency of an integer is the number of times it appears in the list. If no such a value exist, return -1. \\
    \indent\qquad Examples: \\
    \indent\qquad search([4, 1, 2, 2, 3, 1]) == 2 \\
    \indent\qquad search([1, 2, 2, 3, 3, 3, 4, 4, 4]) == 3 \\
    \indent\qquad search([5, 5, 4, 4, 4]) == -1 \\
    \indent\qquad 1. Create a frequency dict. \\
    \indent\qquad 2. Sort the input list. \\
    \indent\qquad 3. Loop the input list, if frequency no lesser than the integer, set result. \\
    \indent\qquad 4. Return the result. \\
    \indent\qquad '''

\end{mybox}
\bigskip

\section{Self-planning Prompt for MBPP Benchmarks} \label{self_planning_prompt_for_mbpp}

\begin{mybox}

\noindent Write a function to sum the length of the names of a given list of names after removing the names that start with a lowercase letter. \\
1. Loop the input list. \\
2. If the name not start with lowercase letter, add the length of the name to result. \\
3. Return the result.     

\noindent \raisebox{1ex}{\rule{\linewidth}{0.2mm}}

\noindent Write a function to increment the numeric values in the given strings by k. \\
1. Loop the input list. \\
2. If a string is a number, increment it. \\
3. Return modified list.
    
\noindent \raisebox{1ex}{\rule{\linewidth}{0.2mm}}

\noindent Write a python function to find sum of all prime divisors of a given number. \\
1. Create a inner function to check if a number is prime. \\
2. Loop all number less than the input that is prime. \\
3. Check if the input is divisible by that. \\
4. Return the result. 
    
\noindent \raisebox{1ex}{\rule{\linewidth}{0.2mm}}

\noindent Write a function to find the lateral surface area of a cone. \\
1. Calculate the generatrix of the cone. \\
2. Return the result. \\
3. Please import inside the function.
    
\noindent \raisebox{1ex}{\rule{\linewidth}{0.2mm}}

\noindent Write a function to remove all tuples with all none values in the given tuple list. \\
1. Loop the given tuple list. \\
2. Check if all elements in the tuple are None. \\
3. If not, append the tuple to the result list. \\
4. Return the result. 
    
\noindent \raisebox{1ex}{\rule{\linewidth}{0.2mm}}

\noindent Write a python function to find the last two digits in factorial of a given number. \\
1. Calculate the factorial of the input number. \\
2. Return the last two digits of it. 
    
\noindent \raisebox{1ex}{\rule{\linewidth}{0.2mm}}

\noindent Write a python function to replace multiple occurence of character by single. \\
1. Create a pattern that the input character repeats mulitiple times. \\
2. Replace the pattern in input string with input character. \\
3. Please import inside the function.
    
\noindent \raisebox{1ex}{\rule{\linewidth}{0.2mm}}

\noindent Write a python function to move all zeroes to the end of the given list. \\
1. Count the number of zeros. \\
2. Remove the zeros from the list. \\
3. Append the zeros to the end of the list. \\
4. Return the list. \\

\end{mybox}

\bigskip
\section{Instances of Baseline Prompt} \label{example_baseline}
\smallskip
\noindent \centerline{\textbf{Instance of Chain-of-Thought Prompting with Self-planning Format}}

\begin{mybox}
\noindent def encrypt(s):\\ 
    \indent\qquad '''\\
    \indent\qquad Create a function encrypt that takes a string as an argument and returns a string encrypted with the alphabet being rotated. The alphabet should be rotated in a manner such that the letters shift down by two multiplied to two places.\\
    \indent\qquad For example:\\
    \indent\qquad encrypt('hi') returns 'lm'\\
    \indent\qquad encrypt('asdfghjkl') returns 'ewhjklnop'\\
    \indent\qquad encrypt('gf') returns 'kj'\\
    \indent\qquad encrypt('et') returns 'ix'\\
    \indent\qquad Let's think step by step.\\
    \indent\qquad 1. Create a string "alphabet" with all letters of the alphabet.\\
    \indent\qquad 2. Assign the number of places to shift the letters to a variable "bias".\\
    \indent\qquad 3. Initialize a string "result" with an empty string.\\
    \indent\qquad 4. Iterate over the characters of the string "s".\\
    \indent\qquad 5. Find the index of the character in the string "alphabet".\\
    \indent\qquad 6. Add the number of places to shift the letters to the index.\\
    \indent\qquad 7. If the index is larger than 25, subtract 26 from the index.\\
    \indent\qquad 8. Add the character at the index to the string "result".\\
    \indent\qquad 9. Return the string "result".\\
    \indent\qquad '''
\end{mybox}
\bigskip

\noindent \centerline{\textbf{Instance of Extremely Concise Style Self-planning Prompt}}

\begin{mybox}
\noindent def encrypt(s):\\ 
    \indent\qquad '''\\
    \indent\qquad Create a function encrypt that takes a string as an argument and returns a string encrypted with the alphabet being rotated. The alphabet should be rotated in a manner such that the letters shift down by two multiplied to two places.\\
    \indent\qquad For example:\\
    \indent\qquad encrypt('hi') returns 'lm'\\
    \indent\qquad encrypt('asdfghjkl') returns 'ewhjklnop'\\
    \indent\qquad encrypt('gf') returns 'kj'\\
    \indent\qquad encrypt('et') returns 'ix'\\
    \indent\qquad 1. Alphabet, bias 4.\\
    \indent\qquad 2. Latter bias, append.\\
    \indent\qquad '''
\end{mybox}
\bigskip

\centerline{\textbf{Instance of Ground-truth Planning Prompt}}

\begin{mybox}
\noindent def encrypt(s):\\
    \indent\qquad '''\\
    \indent\qquad Create a function encrypt that takes a string as an argument and returns a string encrypted with the alphabet being rotated. The alphabet should be rotated in a manner such that the letters shift down by two multiplied to two places.\\
    \indent\qquad '''\\
    \indent\qquad alphabet = 'abcdefghijklmnopqrstuvwxyz'\\
    \indent\qquad bias = 2 * 2\\
    \indent\qquad result = ''\\
    \indent\qquad for char in s:\\
    \indent\qquad\qquad index = alphabet.find(char) + bias\\
    \indent\qquad\qquad if index $>$ 25:\\
    \indent\qquad\qquad\qquad index = index - 26\\
    \indent\qquad\qquad result += alphabet[index]\\
    \indent\qquad return result\\
    \indent Write steps according to the code.\\
    \indent\qquad 1. Create a alphabet, bias two places multiplied by two.\\
    \indent\qquad 2. Loop the input, find the latter bias letter in alphabet.\\
    \indent\qquad 3. Return result.
\end{mybox}

\bigskip
\centerline{\textbf{Instance of Self-planning Prompt (One-phase)}}

\begin{mybox}
\noindent def encrypt(s):\\ 
    \indent\qquad '''\\
    \indent\qquad Create a function encrypt that takes a string as an argument and returns a string encrypted with the alphabet being rotated. The alphabet should be rotated in a manner such that the letters shift down by two multiplied to two places.\\
    \indent\qquad For example:\\
    \indent\qquad encrypt('hi') returns 'lm'\\
    \indent\qquad encrypt('asdfghjkl') returns 'ewhjklnop'\\
    \indent\qquad encrypt('gf') returns 'kj'\\
    \indent\qquad encrypt('et') returns 'ix'\\
    \indent\qquad 1. Create a alphabet, bias two places multiplied by two.\\
    \indent\qquad 2. Loop the input, find the latter bias letter in alphabet.\\
    \indent\qquad 3. Return result.\\
    \indent\qquad '''\\
    \indent\qquad $\#$ Write your code here.\\
    \indent\qquad alphabet = 'abcdefghijklmnopqrstuvwxyz'\\
    \indent\qquad bias = 2 * 2\\
    \indent\qquad result = ''\\
    \indent\qquad for char in s:\\
    \indent\qquad\qquad index = alphabet.find(char) + bias\\
    \indent\qquad\qquad if index $>$ 25:\\
    \indent\qquad\qquad\qquad index = index - 26\\
    \indent\qquad\qquad result += alphabet[index]\\
    \indent\qquad return result
\end{mybox}

\newpage
\section{Human Evaluation Questionnaire} \label{questionnaire}
\begin{figure*}[htbp]
\centering
\begin{minipage}[th!]{0.49\textwidth}
\centering
\begin{adjustbox}{frame}
\includegraphics[width=7cm]{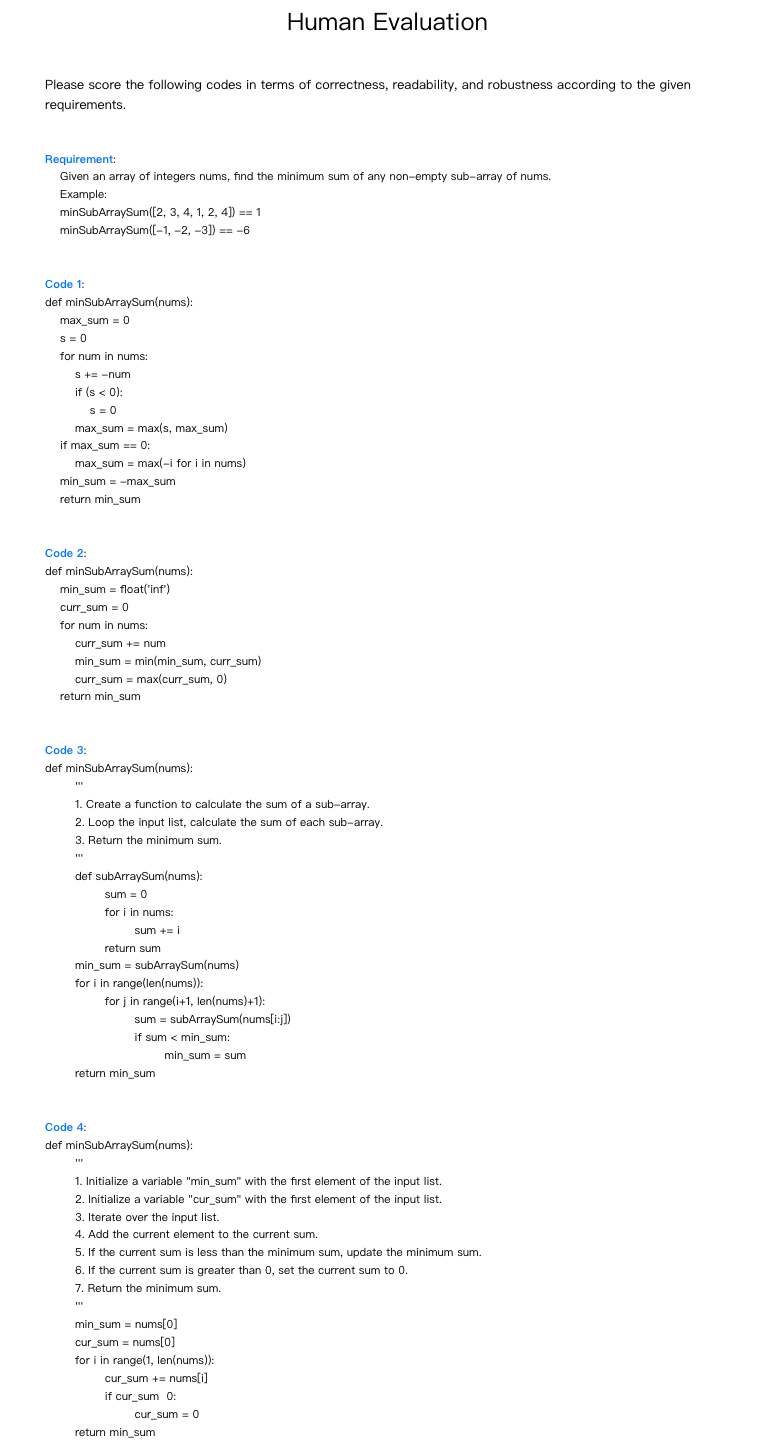}
\end{adjustbox}
\caption{Questionnaire Part 1} 
\label{questionnaire_1}
\end{minipage}
\begin{minipage}[th!]{0.49\textwidth}
\centering
\begin{adjustbox}{frame}
\includegraphics[width=6.8cm]{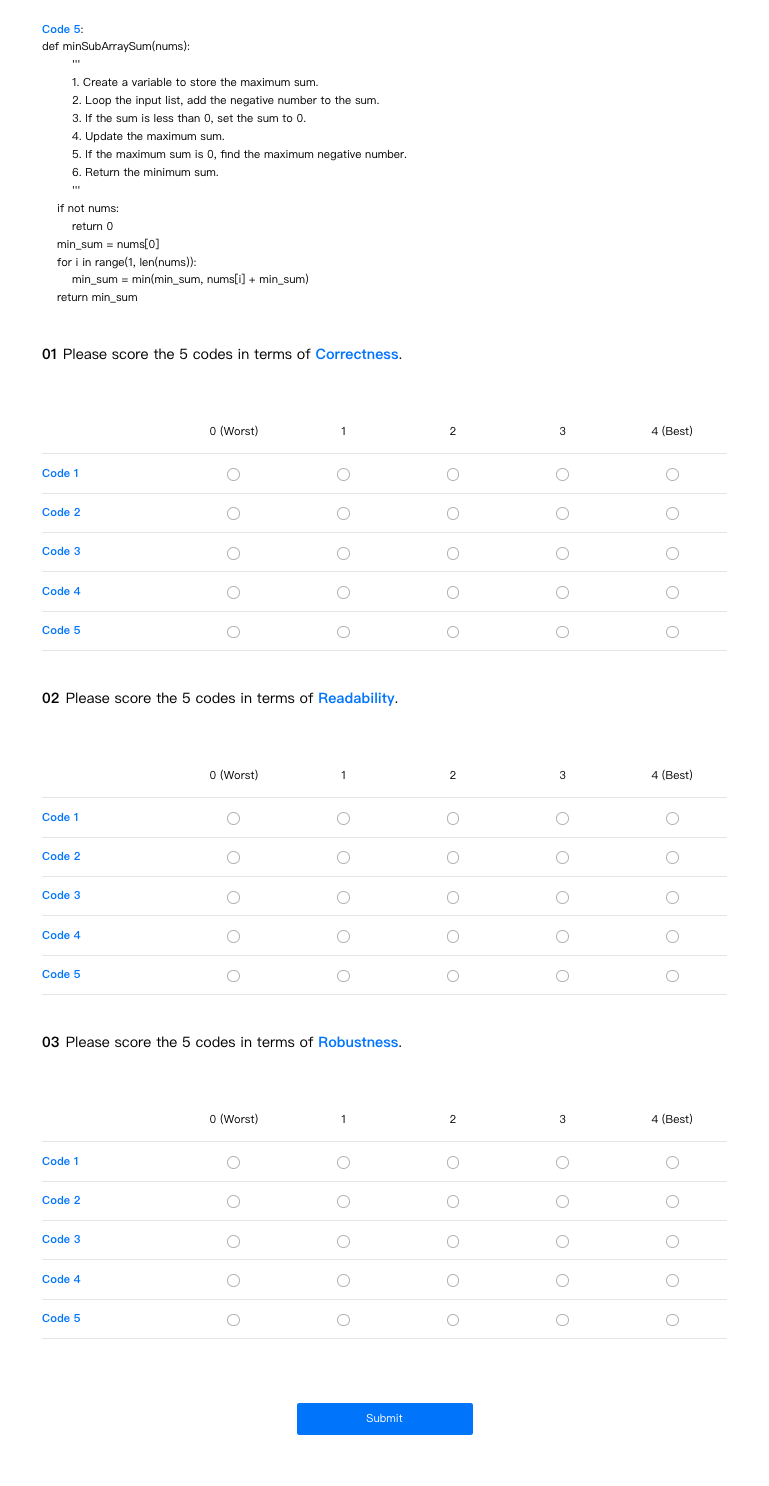}
\end{adjustbox}
\caption{Questionnaire Part 2}
\label{questionnaire_2}
\end{minipage}
\end{figure*}

The questionnaires for the human evaluation of a task are shown in Fig. \ref{questionnaire_1} and Fig. \ref{questionnaire_2}. We distributed 50 questionnaires, which correspond to 50 tasks.

\end{document}